\documentclass[showpacs,epsfig,preprintnumbers,amsmath,amssymb,nofootinbib]{revtex4-1}
\usepackage{graphicx}
\usepackage{dcolumn}
\usepackage{bm}
\usepackage{subfigure}
\usepackage{comment}
\usepackage{epstopdf}
\begin{document}
\def\d{{\rm d}}
\def\Epos{E_{\rm pos}}
\def\ap{\approx}
\def\eff{{\rm eft}}
\def\L{{\cal L}}
\newcommand{\vev}[1]{\langle {#1}\rangle}
\newcommand{\CL}   {C.L.}
\newcommand{\dof}  {d.o.f.}
\newcommand{\eVq}  {\text{EA}^2}
\newcommand{\Sol}  {\textsc{sol}}
\newcommand{\SlKm} {\textsc{sol+kam}}
\newcommand{\Atm}  {\textsc{atm}}
\newcommand{\Chooz}{\textsc{chooz}}
\newcommand{\Dms}  {\Delta m^2_\Sol}
\newcommand{\Dma}  {\Delta m^2_\Atm}
\newcommand{\Dcq}  {\Delta\chi^2}
\newcommand{\nbb}{$\beta\beta_{0\nu}$ }
\def\VEV#1{\left\langle #1\right\rangle}
\let\vev\VEV
\def\e6{E(6)}
\def\10{SO(10)}
\def\21{SA(2) $\otimes$ U(1) }
\def\321{$\mathrm{SU(3) \otimes SU(2) \otimes U(1)}$ }
\def\lr{SA(2)$_L \otimes$ SA(2)$_R \otimes$ U(1)}
\def\422{SA(4) $\otimes$ SA(2) $\otimes$ SA(2)}
\newcommand{\AHEP}{%
School of physics, Institute for Research in Fundamental Sciences
(IPM)\\P.O.Box 19395-5531, Tehran, Iran\\

  }
\newcommand{\Tehran}{%
School of physics, Institute for Research in Fundamental Sciences (IPM)
\\
P.O.Box 19395-5531, Tehran, Iran}
\def\roughly#1{\mathrel{\raise.3ex\hbox{$#1$\kern-.75em
      \lower1ex\hbox{$\sim$}}}} \def\lsim{\roughly<}
\def\gsim{\roughly>}
\def\ltap{\raisebox{-.4ex}{\rlap{$\sim$}} \raisebox{.4ex}{$<$}}
\def\gtap{\raisebox{-.4ex}{\rlap{$\sim$}} \raisebox{.4ex}{$>$}}
\def\lsim{\raise0.3ex\hbox{$\;<$\kern-0.75em\raise-1.1ex\hbox{$\sim\;$}}}
\def\gsim{\raise0.3ex\hbox{$\;>$\kern-0.75em\raise-1.1ex\hbox{$\sim\;$}}}



\title{Measuring Dirac CP-violating phase with intermediate energy beta beam facility}

\date{\today}
\author{P. Bakhti}\email{pouya\_bakhti@ipm.ir}
\author{Y. Farzan}\email{yasaman@theory.ipm.ac.ir}
\affiliation{\Tehran}
\begin{abstract}
Taking the established nonzero value of $\theta_{13}$, we study
the possibility of extracting the Dirac CP-violating phase by a
beta beam facility with a boost factor $100<\gamma<450$. We
compare the performance of different setups with different
baselines, boost factors and detector technologies. We find that
an antineutrino beam from  $^6$He decay with a baseline of
$L=1300$ km  has a very promising CP discovery
potential using a 500 kton Water Cherenkov (WC) detector.
Fortunately this baseline corresponds to the distance between
FermiLAB to Sanford underground research facility in South Dakota.

\end{abstract}
{}
\date{\today}
\maketitle
\section{Introduction}

The developments in neutrino physics in recent 15 years have been
overwhelmingly fast. Nonzero neutrino mass has been established
and five out of nine neutrino mass parameters have been measured
with remarkable precision. In 2012, at last the relatively small
mixing angle, $\theta_{13}$ was measured
\cite{An:2012eh,Ahn:2012nd,Abe:2011fz}. This nonzero value of
$\theta_{13}$ opens up the possibility of having CP-violating
effects in the neutrino oscillations; {\it i.e.,} $P(\nu_\alpha
\to \nu_\beta)\ne P(\bar{\nu}_\alpha \to \bar{\nu}_\beta)$. With
this nonzero value of $\theta_{13}$, the quest for measuring the
Dirac CP-violating phase, $\delta_D$, has been gathering momentum.
A  well-studied way to extract $\delta_D$ is
the precision measurement and comparison of $P(\nu_\mu \to \nu_e)$
and
 $P(\bar{\nu}_\mu \to \bar{\nu}_e)$ by superbeam  and neutrino factory facilities
 \cite{superbeams}.  However, this is not the
 only way. In fact by studying the energy dependence of just one
 appearance mode {\it e.g.,} $P(\nu_\mu \to \nu_e)$, the value of
 $\delta_{CP}$ can be extracted \cite{WB}.
 In \cite{Farzan:2002ct}, a novel method for extracting $\delta_D$
 (or more precisely $\cos \delta_D$) was suggested that was based
 on reconstructing the unitary triangle in the lepton sector.
  The idea of reconstructing the unitary triangle in the
 lepton sector has been later on studied in \cite{triangles}.

Recently, beta beam facilities producing $\nu_e$ or $\bar{\nu}_e$
beams  from
 the decay of relativistic ions \cite{Zucchelli} have been  proposed and studied as an alternative
  machine to
 establish CP-violation in the neutrino sector. Most studies were however performed before the measurement of
 $\theta_{13}$, with a focus on the CP-discovery reach for values of $\theta_{13}$
 much smaller than the measured one
\cite{Sanjib}. Ref. \cite{Orme} shows that using the
 $\nu_e$ beam from $^{18}$Ne decay with energies peaked around
 1.5-2 GeV, information on $\delta_D$ can be extracted without a
 need for an antineutrino beam. In this setup, the boost factor of
 the decaying ions is $\gamma=450$. In the present work, we shall
 consider a similar setup; however, with  lower boost factors
 yielding neutrino energies below the 13-resonance energy in the
 mantle which is about 6.5 GeV \cite{Agarwalla:2013tza}.
 For a detailed analysis of the  matter effects see \cite{blennow}.
 For a neutrino beam with a given energy in the range
 400~MeV$<E_\nu<$1.5~GeV, the oscillation probability can be approximately
 written as
 \begin{equation} \label{Pemu} P_{e\mu}\simeq |c_{12}^mc_{23}(e^{i\lambda_2}-e^{i\lambda_1})+s_{13}^ms_{23}
 e^{-i\delta_D}(e^{i \lambda_3}-e^{i \lambda_2})|^2\end{equation}
where $c_{12}^m\ll 1$ is the cosine of the 12-mixing angle in matter and
$\lambda_i$ are the phases resulting from the propagation; {\it
i.e.,} for a constant density $\lambda_i=(m_i^2)_{eff}L/(2E)$.
For the antineutrino mode, a similar equation holds with $s_{12}^m\ll 1$ and
\begin{equation}
\label{PBARemu} P_{\bar{e}\bar{\mu}}\simeq |s_{12}^mc_{23}(e^{i\lambda_2}-e^{i\lambda_1})+s_{13}^ms_{23}
 e^{i\delta_D}(e^{i \lambda_3}-e^{i \lambda_1})|^2\ .\end{equation}
Notice that in the above formulas, the deviations of the values of
$\theta_{23}$ and $\delta_D$ in matter from those in vacuum  are
neglected. These deviations are of order of $\Delta m_{12}^2/\Delta
m_{13}^2$ \cite{blennow}. As long as $|\lambda_2-\lambda_1|\sim 1$,
the two terms in Eq. (\ref{Pemu}) as well as those in Eq.
(\ref{PBARemu}) are of the same order so the interference terms
which are sensitive to $\delta_D$ are of order of the oscillation
probabilities themselves. This means that the variation in the
oscillation probabilities due to the change of $\delta_D$ within $(0
\ \pi)$ is of order of the oscillation  probabilities, themselves.
As a result for these energies and
 $|\lambda_2-\lambda_1|\sim 1$, even a moderate precision in the measurement of the probabilities will be enough to extract the value of $\delta_D$.

The flux at the detector scales as $\gamma^2$ and the scattering
cross section of neutrinos increases by increasing the energy
({\it i.e.,} increasing $\gamma$). As a result, for a given
baseline, the statistics increases with $\gamma$. Based on this
observation, most attention in the recent years has been given to
$\gamma>300$. However, one should bear in mind that for
$\gamma<300$, there is the advantage of using very large Water
Cherenkov (WC) detectors. In this energy range, the neutrino
interaction will be dominantly quasi-elastic and its scattering cross section
is known with high precision.
In the literature, the CP-discovery potential of a beta beam setup
from CERN to Frejus with $\gamma<150$ and $L=130$~km has been
investigated \cite{memphys}. Moreover, varying the values of
$\gamma$ and baselines, it has been shown that for $150<\gamma<300$
with 500 kton WC detector \cite{Huber,Mezzetto:2003ub,economy} or
iron calorimeter \cite{Greenfield}, there is a very good chance of
CP-discovery. Ref.  \cite{Sanjib} explores the
$\theta_{13}-\delta_D$ discovery reach with a 300 kton WC and 50
kton LAr detector at Deep Underground Science and Engineering
Laboratory (DUSEL), taking maximum boosts possible at Tevatron. Now
that the value of $\theta_{13}$ is measured and found to be
sizeable, reconsidering $\gamma<300$ setup is imperative. In vacuum,
the dependence of the oscillation probability on $L$ and $\gamma$
would be through $L/\gamma$. However, for setups under
consideration, because of the matter effects, the dependence on $L$
and $ \gamma$ is more sophisticated
 so the dependence on $E$ and $L$ has to be investigated
separately. In particular, while Ref. \cite{Huber}  focuses on
$L/\gamma=2.6$~km, we have found that for $L/\gamma>2.6$~km, there is a
very good discovery potential. The present paper
is devoted to such a study.

In section II, we describe the inputs and how we carry out the
analysis. We outline the characteristics of the beam and the
detector as well as the sources of the background and the systematic
errors. In section III, we present our results and analyze them. In
section IV, we summarize our conclusions and propose an optimal
setup for the $\delta_D$ measurement.
\section{The Inputs For Our Analysis }
Using the GloBES software \cite{Globes}, we investigate the
CP-discovery potential of a beta beam setup with various baselines
and beam boost factors, $\gamma$.   For the central values of the
neutrino parameters, we have taken the latest values in Ref
\cite{NuFit}.
 The hierarchy can be determined by other experiments
such as PINGU \cite{pingu,WinteratPINGU} or combining PINGU and DAYA
Bay II results \cite{DayaPINGU} so we assume that hierarchy is known
by the time that such a beta beam setup is ready. We  study both
normal and inverted hierarchies. T2K and Nova can also solve the
octant degeneracy and determine whether $\theta_{23}<45^\circ$ or
$\theta_{23}>45^\circ$ \cite{octant}. The data already excludes the
$\theta_{23}>45^\circ$ solution at 1 $\sigma$ C.L.  For the
uncertainty of the mixing parameters, we take the values that will
be achievable by forthcoming experiments. Namely, we take the
following
 uncertainties: 0.4\% for  $\theta_{12}$  \cite{Capozzi:2013psa},
 1.8\% for
  $\theta_{13}$  \cite{Tortola:2012te},  2\% for $\theta_{23}$  \cite{Raut:2012dm},  0.2\% for $\Delta m_{12}^{2}$
   \cite{Capozzi:2013psa} and 0.7\% for $\Delta m_{13}^{2}$  \cite{pingu}.
    As predefined by GLoBES, we use the matter profiles in \cite{prem}.
We consider 5\% error for matter density. The uncertainties are
treated by the so-called pull-method
 \cite{Globes}.
 While the effects of uncertainty in matter density is
more important for larger baselines, the uncertainties of neutrino
parameters affect  the results from smaller baselines more.

As the source of neutrino (antineutrino) beam, we  take  decays of
$^{18}$Ne ($^6$He):
$$^{18}{\rm Ne}\to ^{18}{\rm F}+e^++\nu_e$$
and
$$^{6}{\rm He}\to ^{6}{\rm Li}+e^-+\bar{\nu}_e.$$
The endpoint energies of these two decays are very close to each
other: $E_0=3.4$~MeV for $^{18}$Ne and $E_0=3.5$~MeV for $^{6}$He.
As a result for equal $\gamma$, the energy spectrum of $\nu_e$ and
$\bar{\nu}_e$ from their decays will be approximately similar. The
pair of $^8$Li and $^8$B isotopes have also been  discussed in the
literature as a potential source of the $\nu_e$ and $\bar{\nu}_e$
beams. In these cases, the endpoints are higher so to have neutrino
beams with energies $E_\nu<1.5$ GeV, the values of $\gamma$ should
be lower than in the case of $^{18}$Ne/$^{6}$He. On the other hand,
the flux at the detector drops as $\gamma^{-2}$ so with the $^8$Li
and $^8$B isotopes, the number of decays should be larger to
compensate for the $\gamma^{-2}$ suppression. We will not  consider
the $^8$Li/$^8$B isotopes in this paper and will focus on the
$^{18}$Ne/$^{6}$He pair.
 For neutrino (antineutrino) mode, we  take $2.2\times 10^{18}$ ($5.8\times
10^{18}$) decays of $^{18}$Ne ($^6$He) per year which seems to be
realistic \cite{Parke}. Tevatron accelerator can accelerate
$^{18}$Ne and $^6$He up to boost factors 586 and 351, respectively.

While the disappearance probabilities (i.e., $P(\nu_e \to \nu_e)$
 or $P(\bar{\nu}_e\to \bar{\nu}_e)$) are not sensitive to
 $\delta_D$, appearance probabilities (i.e., $P(\nu_e \to \nu_\mu)$
 or $P(\bar{\nu}_e\to \bar{\nu}_\mu)$) contain information on
 $\delta_D$. In our analysis, we however employ
both appearance and disappearance modes. In principle, the
disappearance mode can help to reduce the effect of uncertainty in
other parameters but we have found that the effect of turning off
the disappearance mode on the $\delta_D$ measurement is less than 1
\%.  To derive the value of $\delta_D$, the detector
 has to distinguish $\nu_\mu$ from $\nu_e$. We focus on a 500 kton Water
 Cherenkov  (WC) detector and compare its performance with a 50 kton Totaly Active
Scintillator Detector (TASD).

 In the energies of our interest with
$\gamma<300$, the main interaction mode is Charged Current (CC)
quasi-elastic mode with a non-negligible contribution from inelastic
charged current interaction which produce one or more pions along
with the charged lepton. In principle, the quasi-elastic CC events
can be distinguished from the inelastic CC ones by counting the
number of Cherenkov rings. However with a WC detector distinguishing
the two interactions will be challenging.
 We take the signal to be composed of both quasi-elastic and
 inelastic charged current events and conservatively assume WC
 detector cannot distinguish between the two.

 In the case of QE interaction by measuring the energy and the
 direction of the final charged lepton, the energy of the initial
 neutrino can be reconstructed up to an uncertainty of 0.085~GeV
 caused by the Fermi motion of nucleons inside the nucleus. However, in
 the inelastic interaction, a fraction of  the initial energy is
  carried by pions so the energy of the initial neutrino cannot be
 reconstructed by measuring the energy and  the direction of the
 final lepton alone. A WC detector cannot measure the energy
 deposited in hadronic showers so with this technique, the reconstruction of the energy spectrum will be
 possible only for the
 QE interactions. Following the technique in \cite{Huber,29} we take an unknown
 normalization for QE events and use its spectrum as a basis for
 energy reconstruction. Of course with this method, energy reconstruction cannot be carried out on an event by event basis
 and information on the spectrum will be only statistical.
TASD can measure the energy deposited in hadronic showers, too. As a
result, energy reconstruction by TASD can be possible on an event by
event basis.

 As shown in \cite{Agostino}, the background from atmospheric
 neutrinos can be neglected and the main source of background  for
 both TASD and WC detectors are neutral current interactions of the
 beam neutrinos.
In our analysis, the cross sections of {the quasi-elastic, inelastic and neutral current}
interactions we employ the results of \cite{CS}. 
Recently the MiniBooNE collaboration has measured the antineutrino
cross section in the energy range of our interest
\cite{AguilarArevalo:2013hm} with remarkable precision. In the near
future, the measurement of cross section will become even more
precise. Unless otherwise stated, we assume four years of data
taking.

For the treatment of the efficiencies and backgrounds we implement the same methods used in \cite{Huber,29}. While for the purpose of this paper, the methods used in \cite{Huber,29} are adequate, we would like to note that a more complete discussion of reconstruction of events in large WC detectors can be found in \cite{Agostino}. To be more specific similarly to \cite{Huber}, we assume the following
characteristics for the WC detector performance. We take a signal
efficiency of 55\% for neutrinos and that of 75\% for antineutrinos.
We take the uncertainty in normalization of the total signal to be
2.5 \% but as we mentioned above, we take a free normalization for
QE events. We assume a background rejection of 0.3 \% for neutrinos
and 0.25 \% for antineutrinos. The normalization uncertainty of the
background is taken to be 5 \%. For both background and signal, the
calibration error is 0.0001. For the energy reconstruction of the
background, we use the migration matrices tabulated for the GLoBES
package \cite{migration}. We consider the energy range between 0.2
and 3 GeV and divide it into 28 bins. The energy  resolution
 for QE CC interactions is assumed to be of form $0.085
 +0.05\sqrt{E/{\rm GeV}}$~GeV for both muon and electron neutrino
 detection. The first term originates from the Fermi motion of the
 nuleons inside nuclei and the second term reflects the error in
 measuring the energy of the final charged lepton \cite{Sanjib}.

As in Ref. \cite{Huber}, we assume the following features  for TASD:
A signal efficiency of 80\% for $\nu_\mu$ and $\bar\nu_\mu$ and that
of 20\% for $\nu_e$ and $\bar\nu_e$; background rejection of 0.1 \%;
a signal normalization uncertainty of 2.5\%; normalization
uncertainty of 5\%; a calibration error of 0.0001. The energy
resolution is given by $0.03\sqrt{E/{\rm GeV}}$~GeV for muon
(anti)neutrinos and $0.06 \sqrt{E/{\rm GeV}}$ GeV for electron
(anti)neutrinos. The energy range is taken from 0.5 GeV to 3.5 GeV
and divided to  20 bins. We have studied the dependence of our
results on the number of the bins. It seems that the results do not
change by increasing the number of the bins to 30.

\section{Results and the interpretation}
\begin{figure}
\begin{center}
\subfigure[\ Neutrino beam, normal
hierarchy]{\includegraphics[width=0.49\textwidth]{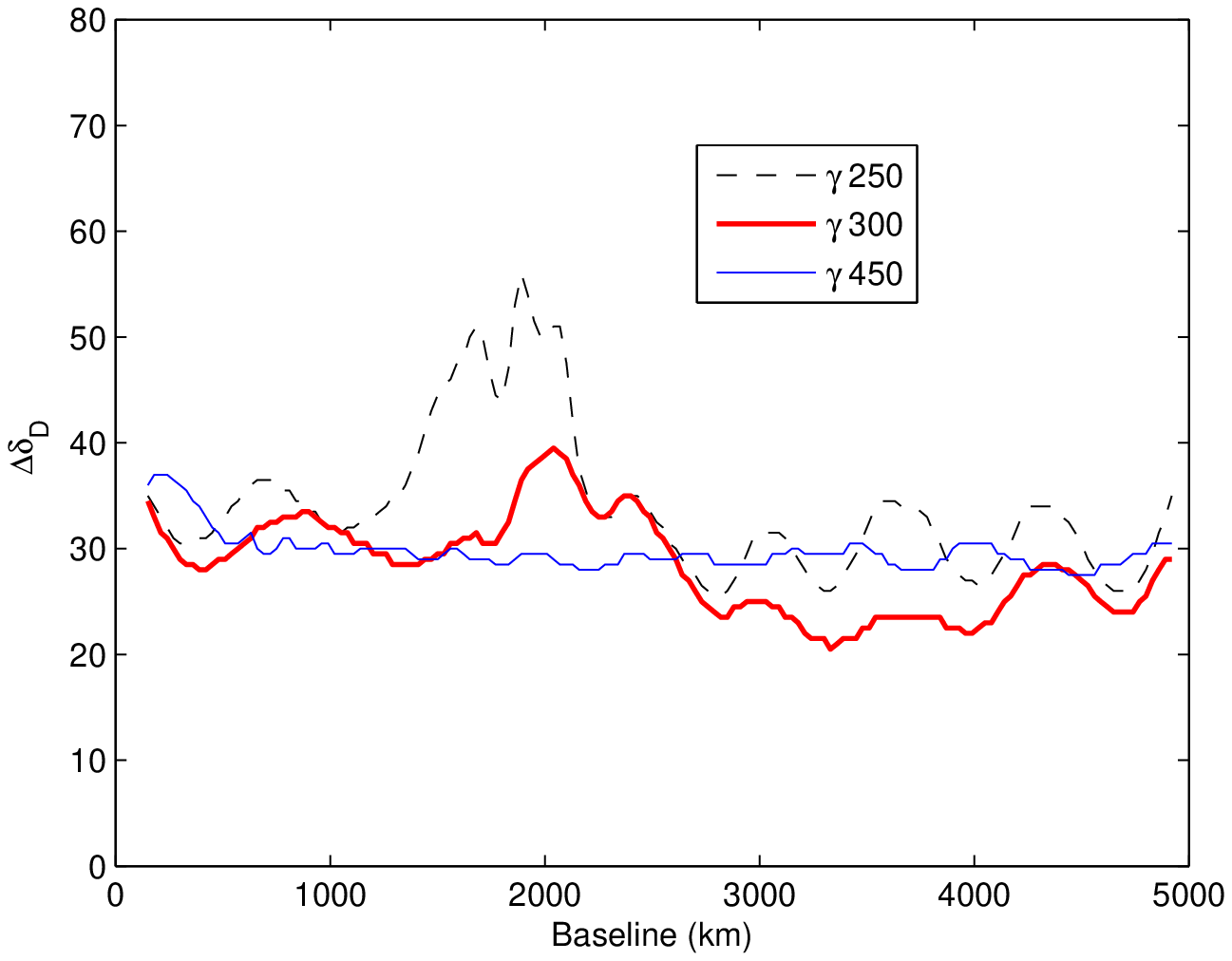}}
\subfigure[\ Neutrino beam, inverted
hierarchy]{\includegraphics[width=0.49\textwidth]{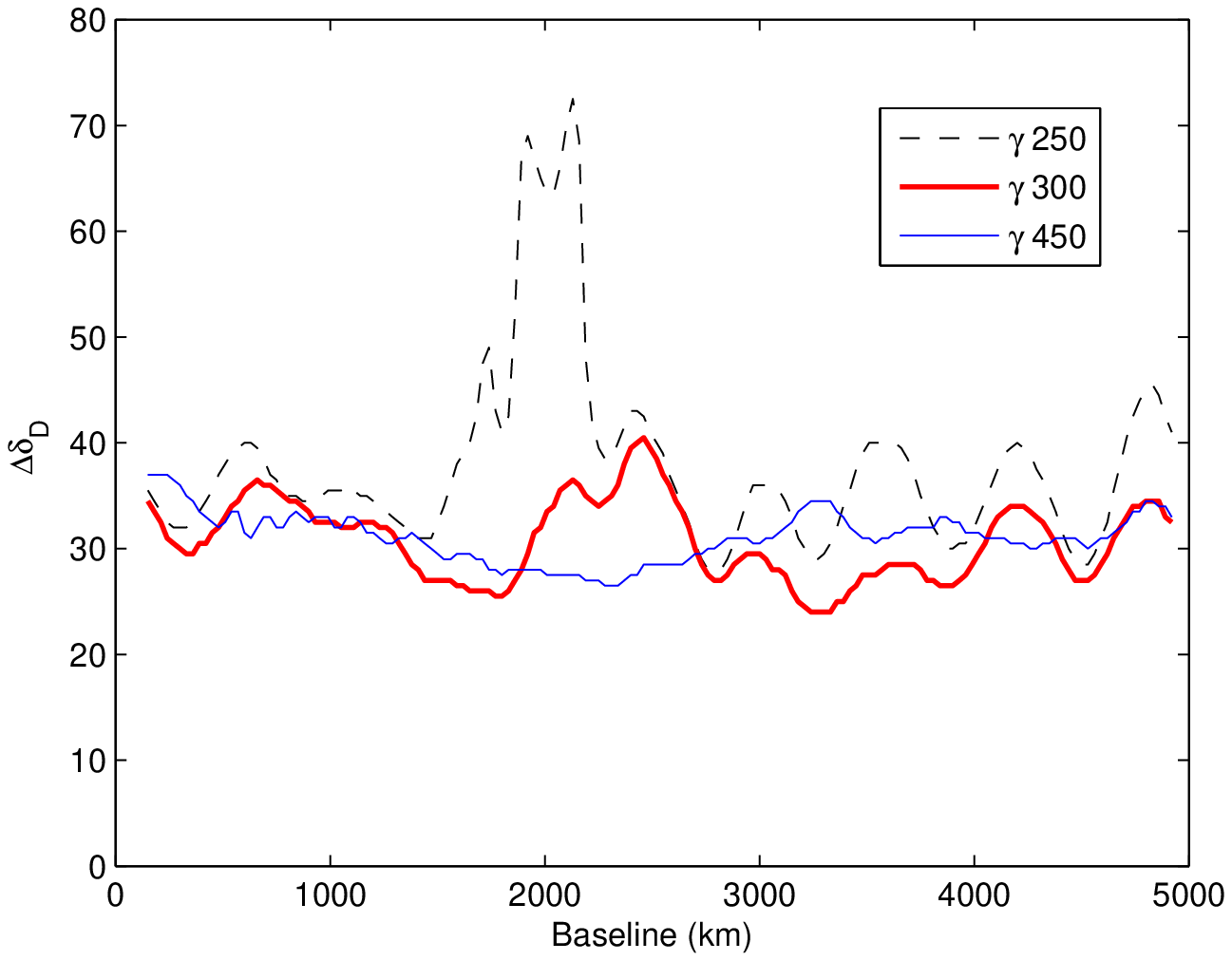}}
\subfigure[\ Antineutrino beam, normal
hierarchy]{\includegraphics[width=0.49\textwidth]{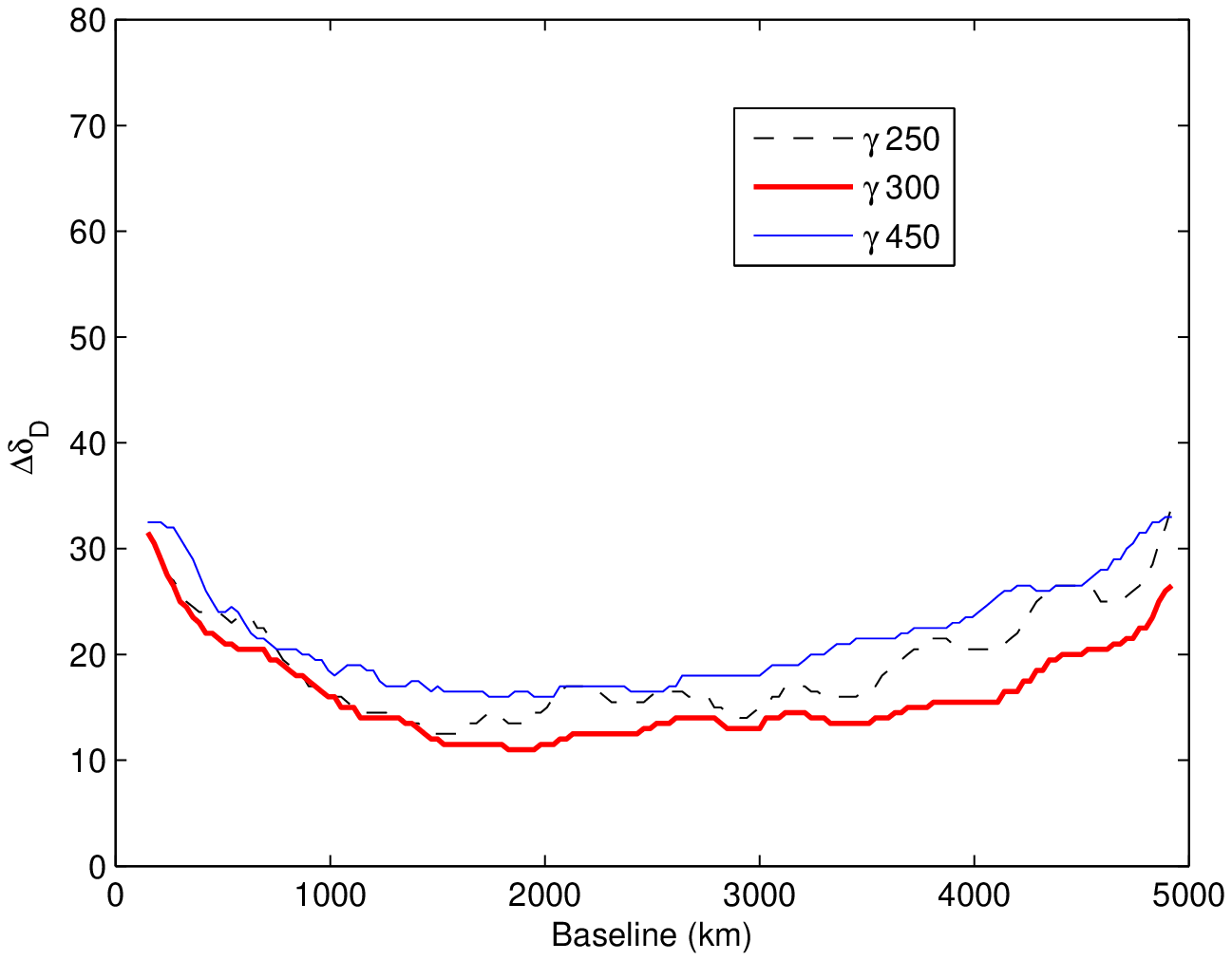}}
\subfigure[\ Antineutrino beam, inverted
hierarchy]{\includegraphics[width=0.49\textwidth]{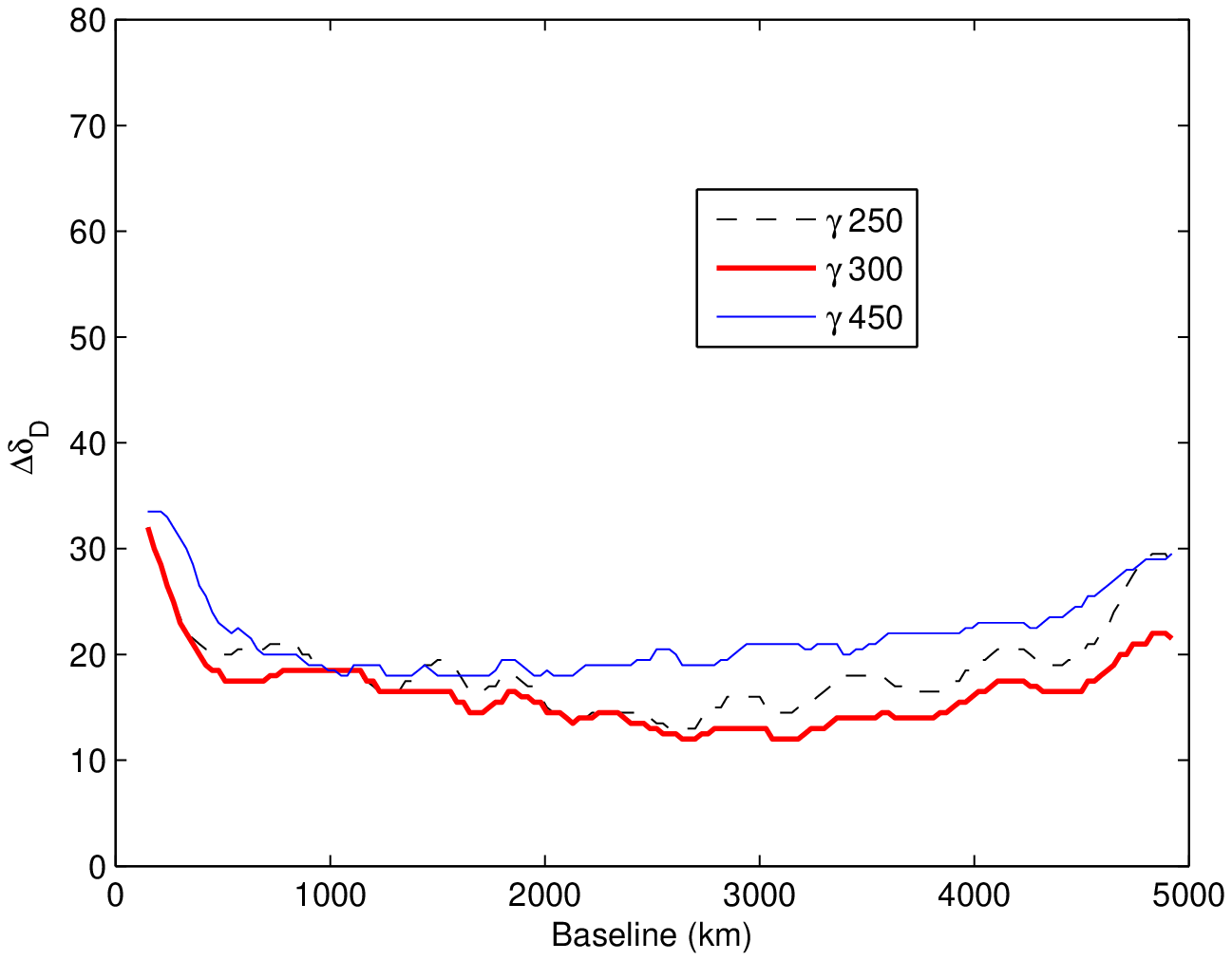}}
\end{center}
\vspace{2cm} \caption[]{Uncertainty within which $\delta_D=90^\circ$
can be measured at 1 $\sigma$ C.L. after four years of data taking
versus baseline for different values of the boost factor. For
$\gamma=450$, a 50 kton TASD detector and for lower $\gamma$, a 500
kton WC detector are assumed.
   In upper (lower)  panels, neutrino (antineutrino) beam with  $2.2 \times 10^{18}$
($5.8 \times 10^{18}$) decays per year is assumed. In left (right)
panels, the hierarchy is taken to be normal (inverted). We have
taken the true values for the left panels as $\Delta
m_{31}^2=2.421\times 10^{-3}$~eV$^2$ (normal hierarchy) and
$\theta_{23}=41.4^\circ$ (first octant). For the right panels we
have taken $\Delta m_{31}^2=-2.35\times 10^{-3}$~eV$^2$ (Inverted
Hierarchy) and the same mixing angles.}

\label{L-dd}

\end{figure}

 In Figs \ref{L-dd} and \ref{nu-and-antiNu-dd}, the vertical axis shows the precision with
which $\delta_D=90^\circ$ can be determined at 1 $\sigma$ \% C.L. We
take $\delta_D=90^\circ$ and define $\Delta \delta_D$ the range for
which $\Delta \chi^2<1$. More precisely,
 $\Delta \delta_D$ is defined as the difference between maximum
 and minimum values of $\delta_D$ around $\delta_D=90^\circ$ for
 which $\Delta \chi^2=1$. From Fig \ref{L-dd}, we observe that
 low energy set-up with $\gamma=300$ and WC detector can
 outperform the setup with $\gamma=450$ and TASD detector for both
 normal and inverted hierarchies. The oscillatory behavior of the
 curves is driven by the 13-splitting and has a frequency given by
 $\sim \Delta m_{31}^2/(2E)$. Such a behavior can be understood by
 the following consideration on Eqs. (\ref{Pemu},\ref{PBARemu}):
 While
$\lambda_2-\lambda_1$  is driven by $(\Delta m_{21}^2)_{eff}$ and
slowly varies with $L$, $\lambda_3-\lambda_2$  is driven by $(\Delta
m_{32}^2)_{eff}$ and varies rapidly. For the values of $L$ that
$\lambda_3-\lambda_2=2n\pi$, the sensitivity is lost. This
consideration explains the oscillatory behavior of Figs. \ref{L-dd}.
Notice, however that this consideration holds for a given $E_\nu$.
If the energy spectrum is wide, the effect will smear out. In other
words, if the number of energy bins from which information on
$\delta_D$ can be deduced ({\it i.e.,} the bins for which the number
of events without oscillation is sizeable and the quasi-elastic
interactions dominate) is relatively large, missing information in
few of these bins for which $(\lambda_3-\lambda_2)\to 0$ will not
affect much the precision in the determination of $\delta_D$. In the
opposite case when at all such bins $(\lambda_3 -\lambda_2)\to 0$,
the precision in $\delta_D$ will be dramatically deteriorated.
Increasing the boost factor increases both the peak energy and the
energy width. Thus, we expect for higher $\gamma$, the oscillatory
behavior to be smeared out. Fig. \ref{L-dd} confirm this
expectation.
In case of antineutrinos, the information on $\delta_D$ can be
deduced from a larger range of spectrum mainly because of the shape
of the spectrum at the source and the fact that for antineutrinos,
the QE interactions dominate over inelastic interaction for a wider
energy range compared to the case of neutrino \cite{CS}. As a
result, the modulation driven by $\Delta m_{31}^2$ is less severe
for antineutrinos.
 As seen in the lower panels of Fig 1, the antineutrino beam with
$\gamma=300$ and a WC detector can achieve an impressive precision
of better than 20$^\circ$ for baselines over 500 km.

\begin{figure}
\begin{center}
\subfigure[\ Neutrino beam, normal
hierarchy]{\includegraphics[width=0.49\textwidth]{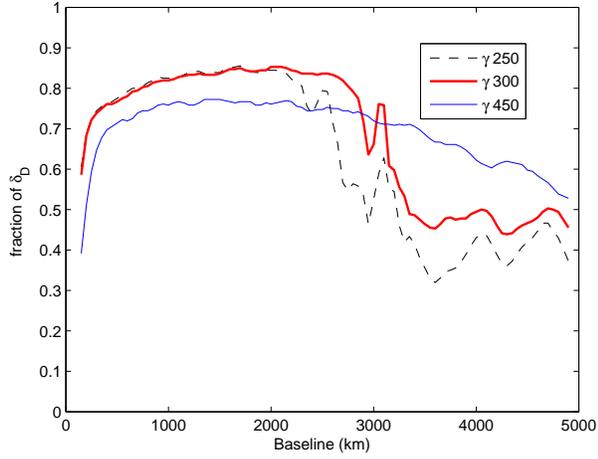}}
\subfigure[\ Neutrino beam, inverted
hierarchy]{\includegraphics[width=0.49\textwidth]{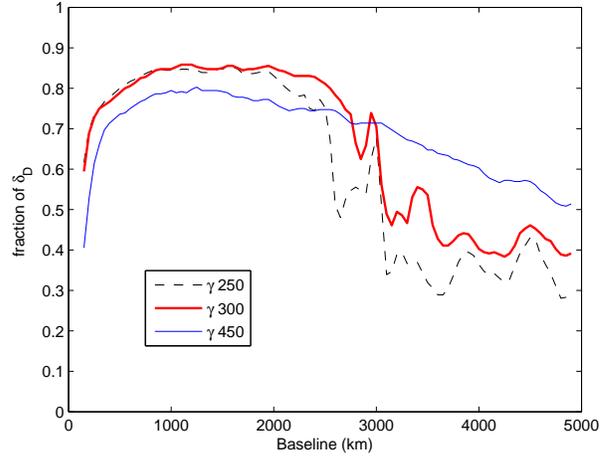}}
\subfigure[\ Antineutrino beam, normal
hierarchy]{\includegraphics[width=0.49\textwidth]{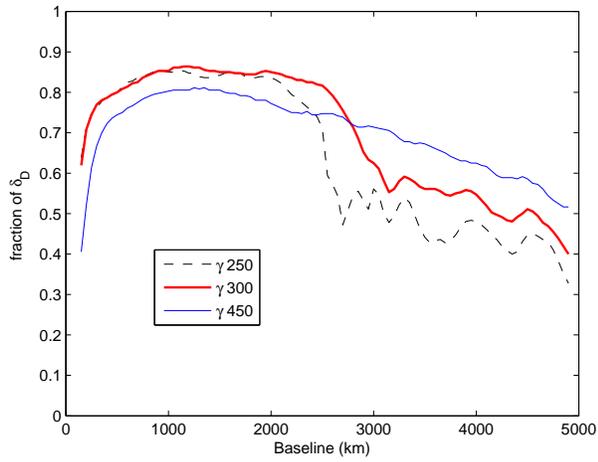}}
\subfigure[\ Antineutrino beam, inverted
hierarchy]{\includegraphics[width=0.49\textwidth]{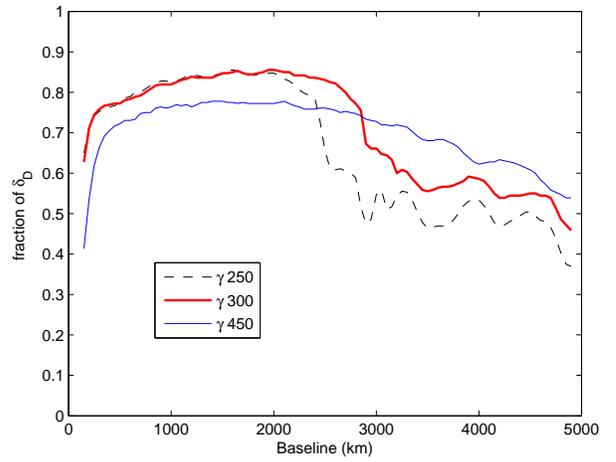}}
\end{center}
\vspace{2cm} \caption[]{The fraction of the $\delta_D$ parameter for
which $CP$ can be established at higher than 95 \% C.L. after four
years of data taking versus baseline. The rest of description is as
in Fig. \ref{L-dd}.}

\label{L-f}

\end{figure}

\begin{figure}
\begin{center}

{\includegraphics[width=1\textwidth]{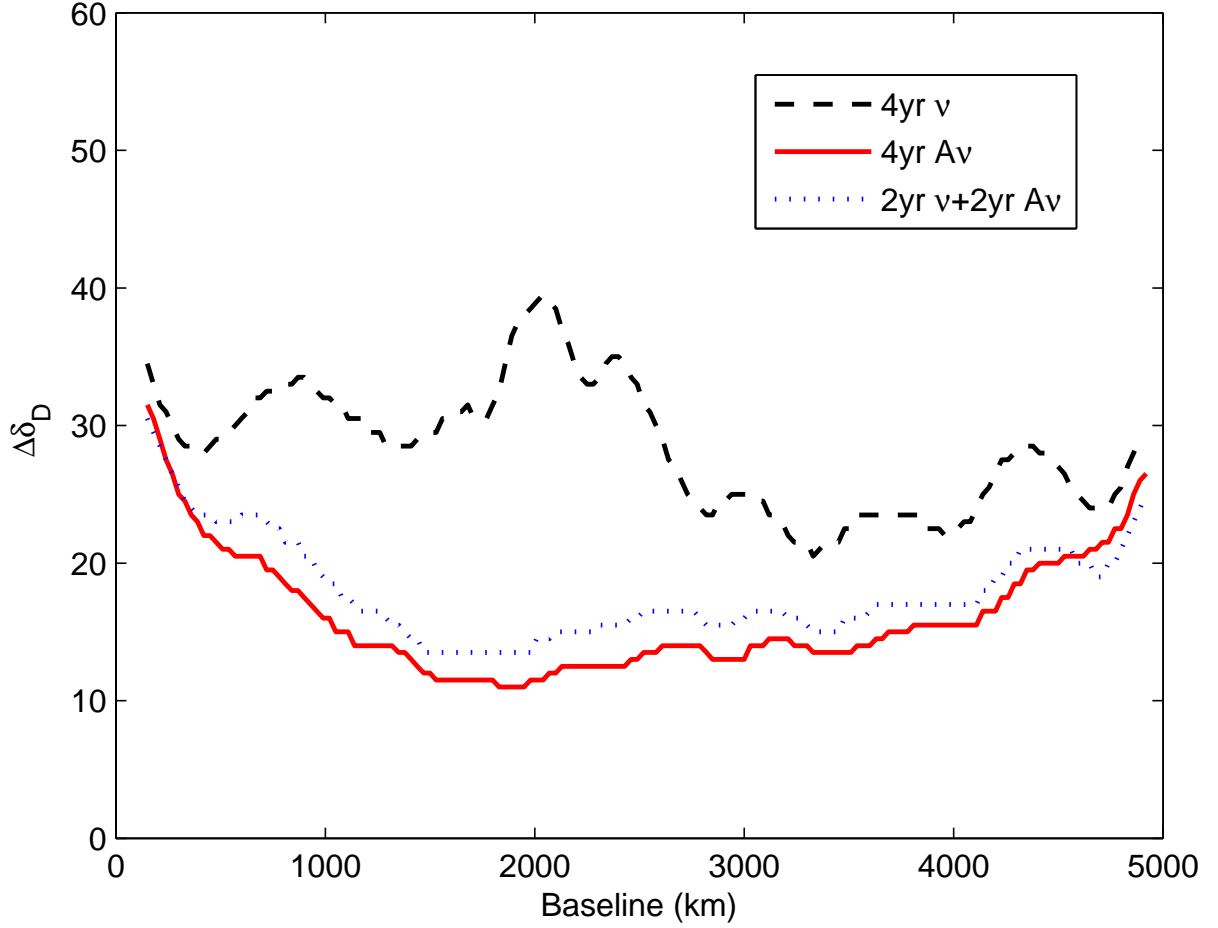}}

\end{center}
\vspace{2cm} \caption[]{Uncertainty within which $\delta_D=90^\circ$
can be measured versus baseline at 1$\sigma$ C.L. The neutrino
parameters are as in Fig.~1. The hierarchy is taken to be normal.
For the neutrino and antineutrino beams, $2.2\times 10^{18}$ and
$5.8\times 10^{18}$ decays per year are assumed, respectively.  The
curves shown with dashed and solid lines respectively show the
results of four years run on neutrino mode from the $^{18}$Ne decay
and  four years run on antineutrino mode from the $^{6}$He decay.
The curve shown by dotted line displays the results of two years of
neutrino run combined with two years of antineutrino run.  The boost
factors of the beams are taken to be 300.}

\label{nu-and-antiNu-dd}

\end{figure}

Figs. \ref{L-f} show the fraction of the parameter $\delta_D$ for
which CP-violation can be established. From these figures, we also
observe that setups with $\gamma<300$ and 500 kton WC detector can
outperform the setup with 50 kton TASD detector and $\gamma=450$ for
$L<2500$ km.

\begin{figure}
\begin{center}

{\includegraphics[width=1\textwidth]{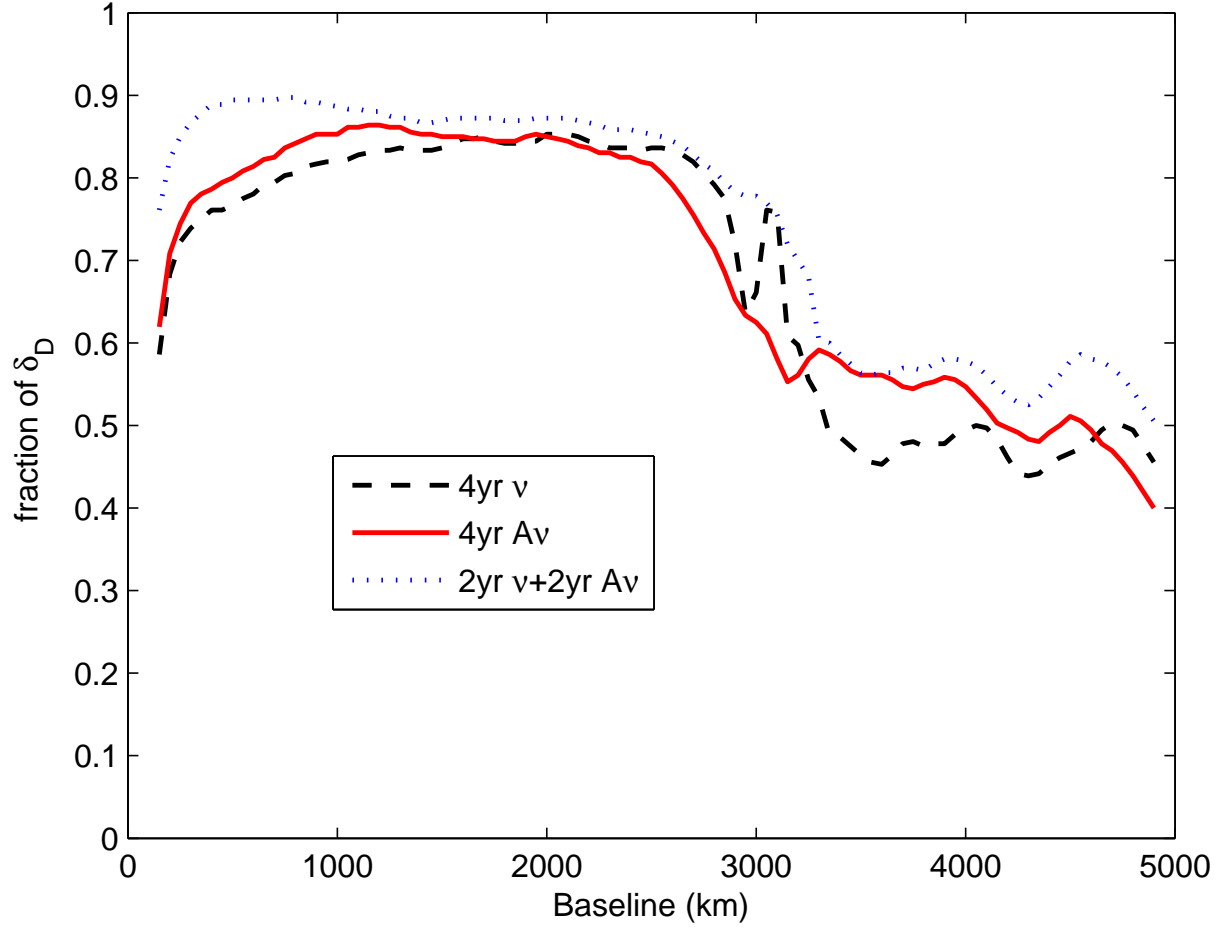}}

\end{center}
\vspace{2cm}
\caption[]{The fraction of the $\delta_D$ parameter for which $CP$
can be established at higher than 95 \% C.L. versus baseline. The rest of description is as that of Fig. \ref{nu-and-antiNu-dd}.}

\label{nu-and-antiNu-f}

\end{figure}

Figs. \ref{nu-and-antiNu-dd} and \ref{nu-and-antiNu-f} compare the
CP-discovery potential of a $\nu$ run with an $\bar{\nu}$ run and a
mixed balanced run. For the antineutrino run the decay rate is taken
to be about 2.6 times that of neutrinos to compensate for the low
cross sections of antineutrinos. For $200~{\rm km}<L<5000$ km, the
antineutrino run seems to outperform both the neutrino run and  the
mixed run in the precision measurement of $\delta_D=90^\circ$. This
result is at odds with the results of \cite{Huber}. However, we
should remember that Ref. \cite{Huber} focuses on a specific value
of $L/\gamma$. In this energy and baseline range, the sensitivity of
average $P(\bar{\nu}_e \to \bar{\nu}_\mu)$ to $\delta$
 is higher than that of average  $P(\nu_e \to \nu_\mu)$.

\begin{figure}
\begin{center}
\subfigure[\ Normal
hierarchy]{\includegraphics[width=0.49\textwidth]{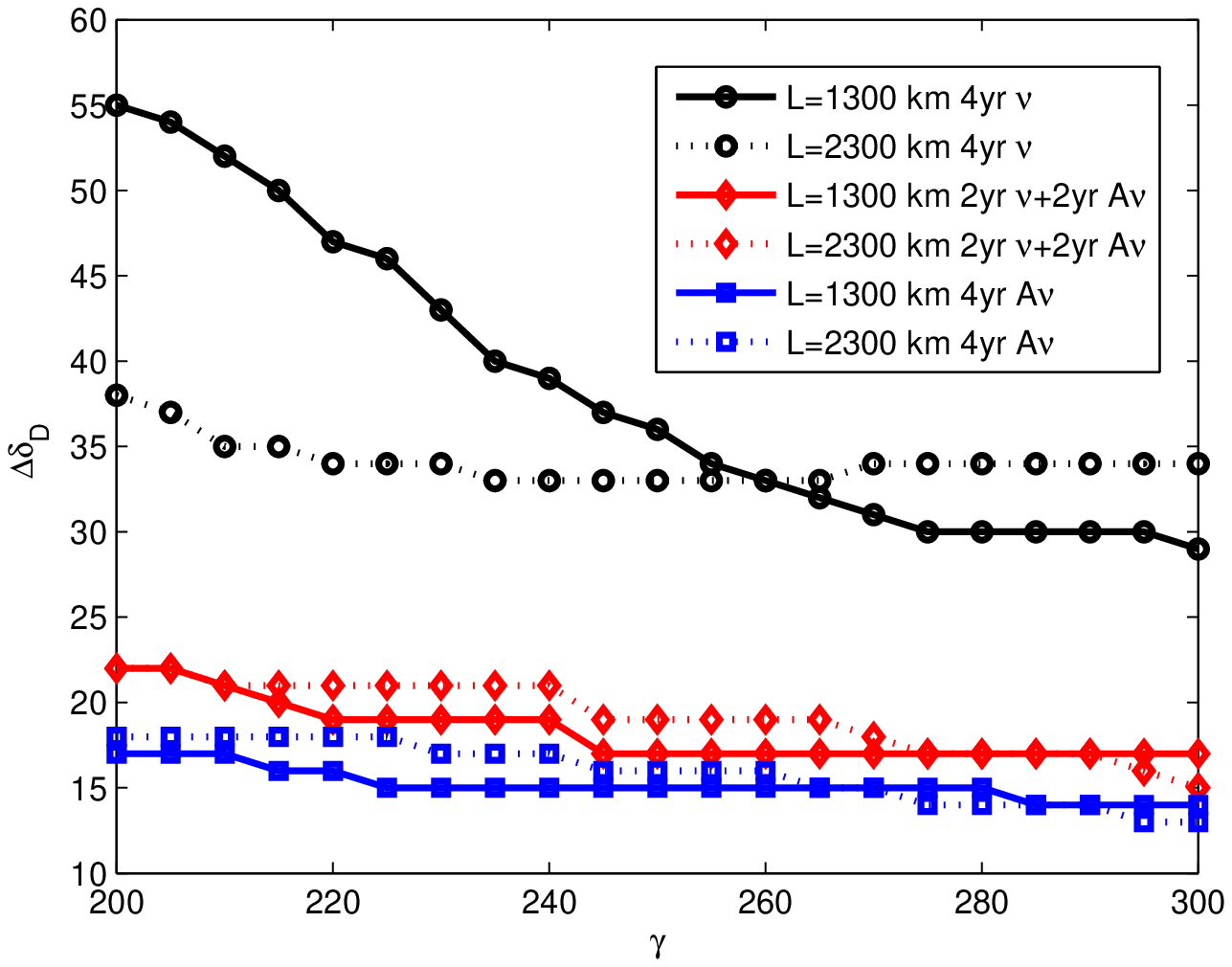}}
\subfigure[\ Inverted
hierarchy]{\includegraphics[width=0.49\textwidth]{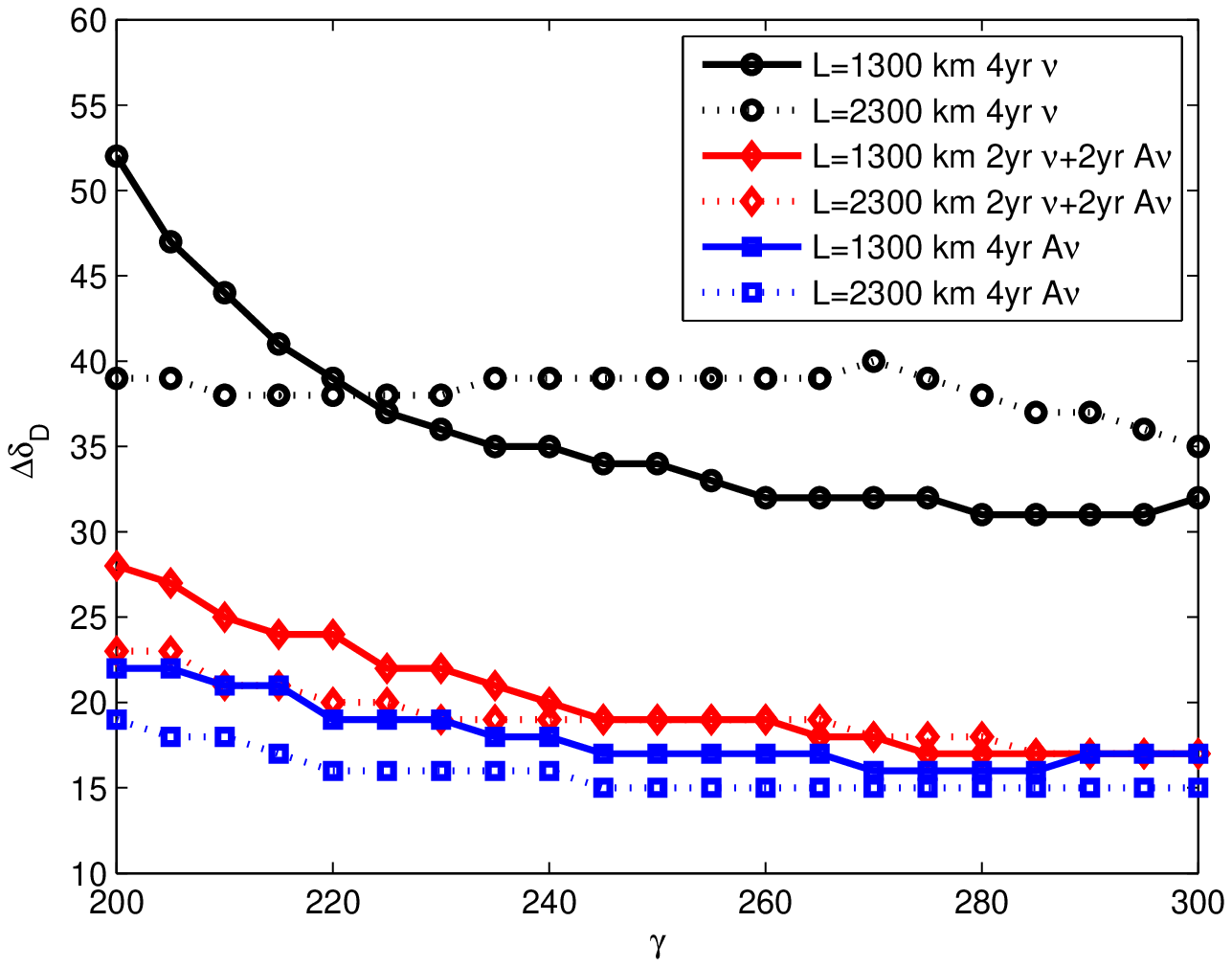}}
\subfigure[\ Normal
hierarchy]{\includegraphics[width=0.49\textwidth]{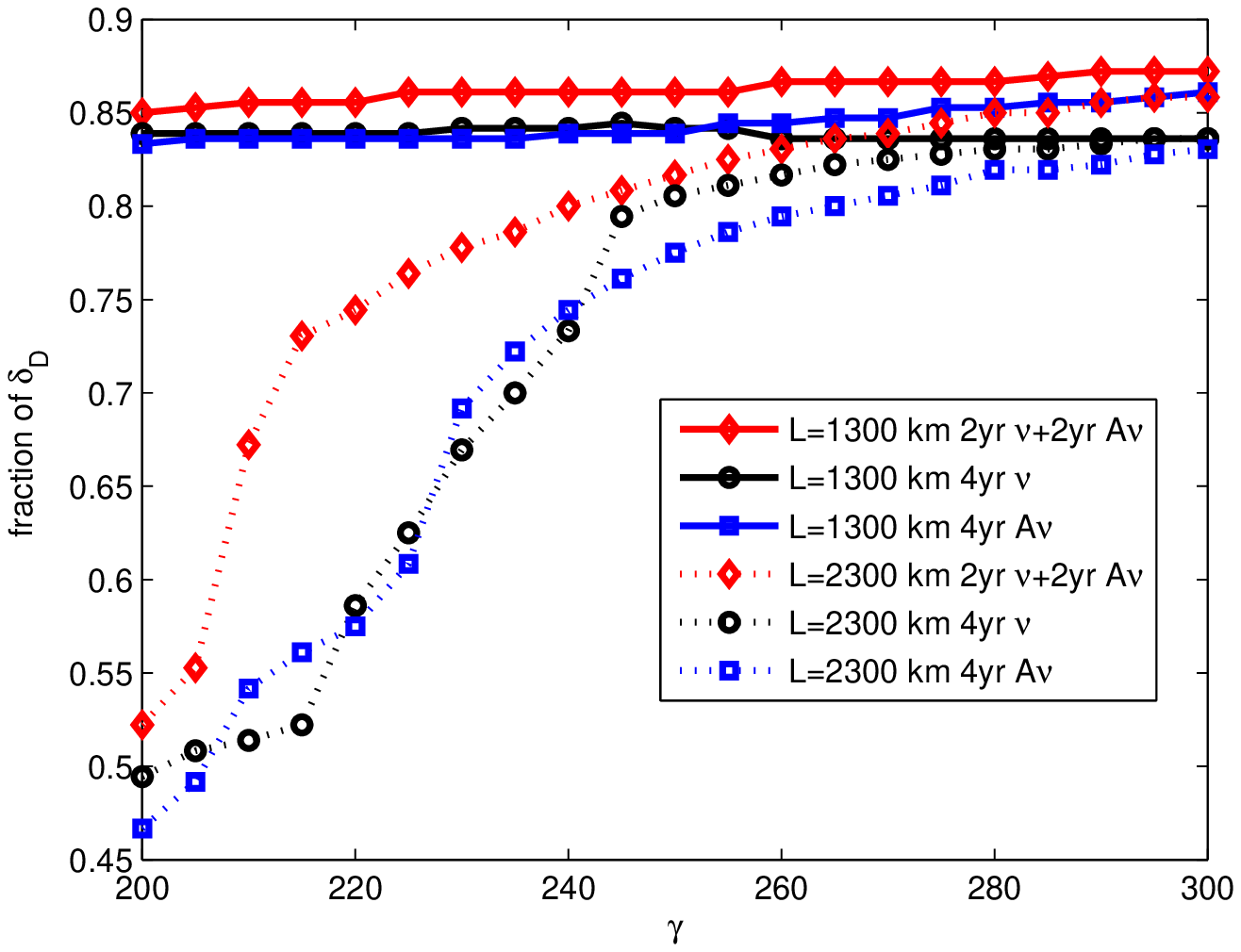}}
\subfigure[\ Inverted
hierarchy]{\includegraphics[width=0.49\textwidth]{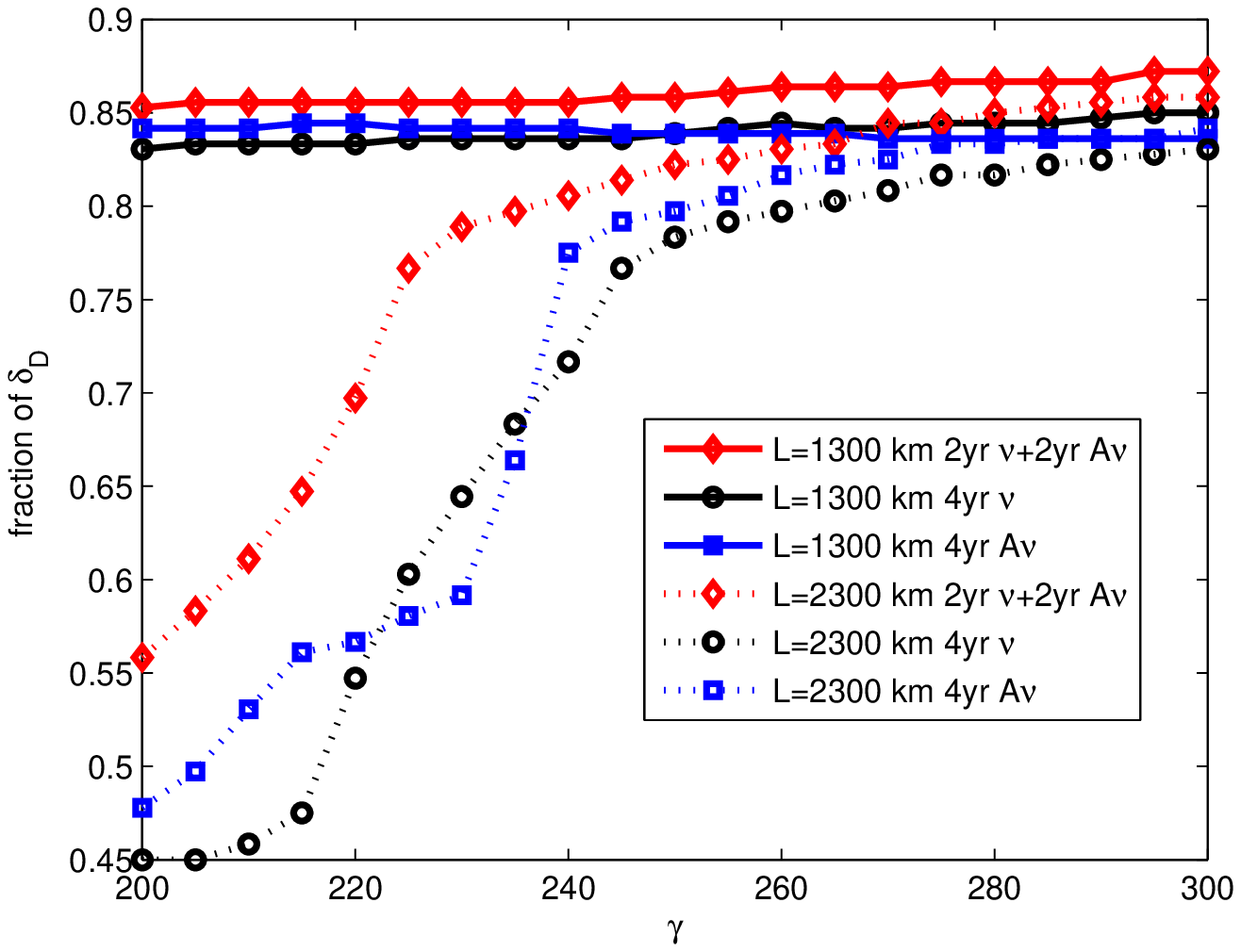}}
\end{center}
\vspace{2cm} \caption[]{The CP-discovery potential by the setups
with baselines  equal to 2300 km  and 1300 km after four years of
data taking versus the boost factors of the neutrino and
antineutrino beams. In the left (right) panels, the hierarchy is
taken to be normal (inverted).  {\it Upper panels}: Uncertainty
within which $\delta_D=90^\circ$ can be measured at 1$\sigma$ C.L.
versus the boost factors of $\nu_e$ and $\bar{\nu}_e$ beams from the
$^{18}$Ne  and $^6$He decay. {\it lower panels: } The fraction of
the $\delta_D$ parameter for which $CP$ can be established at higher
than 95 \% C.L.}

\label{LBNEO}

\end{figure}

 For
$L=1300$ km, (corresponding to the baseline of the LBNE setup from
the FermiLAB to Sanford underground research facility in south
Dakota \cite{LBNE}), we also observe that $\gamma=300$ with WC
detector is promising and can outperform the $\gamma=450$ setup with
TASD detector. Fig. \ref{LBNEO} shows $\Delta \delta_D$ versus
$\gamma$ for  $L=1300$ km  and $L=2300$ km. The latter corresponds
to the baseline for the LBNO setup from CERN to Finland \cite{LBNO}.
Plots show that the setup with $\gamma=200-300$ and $L=1300$ km  can
measure $\delta_D=90^\circ$ with a remarkable precision and  also
have an outstanding coverage of the $\delta_D$ range. At this
baseline, increasing $\gamma$ from 200 to 300 does not much improve
the sensitivity to $\delta_D$.

\begin{figure}
\begin{center}
\subfigure[\ Normal
hierarchy]{\includegraphics[width=0.49\textwidth]{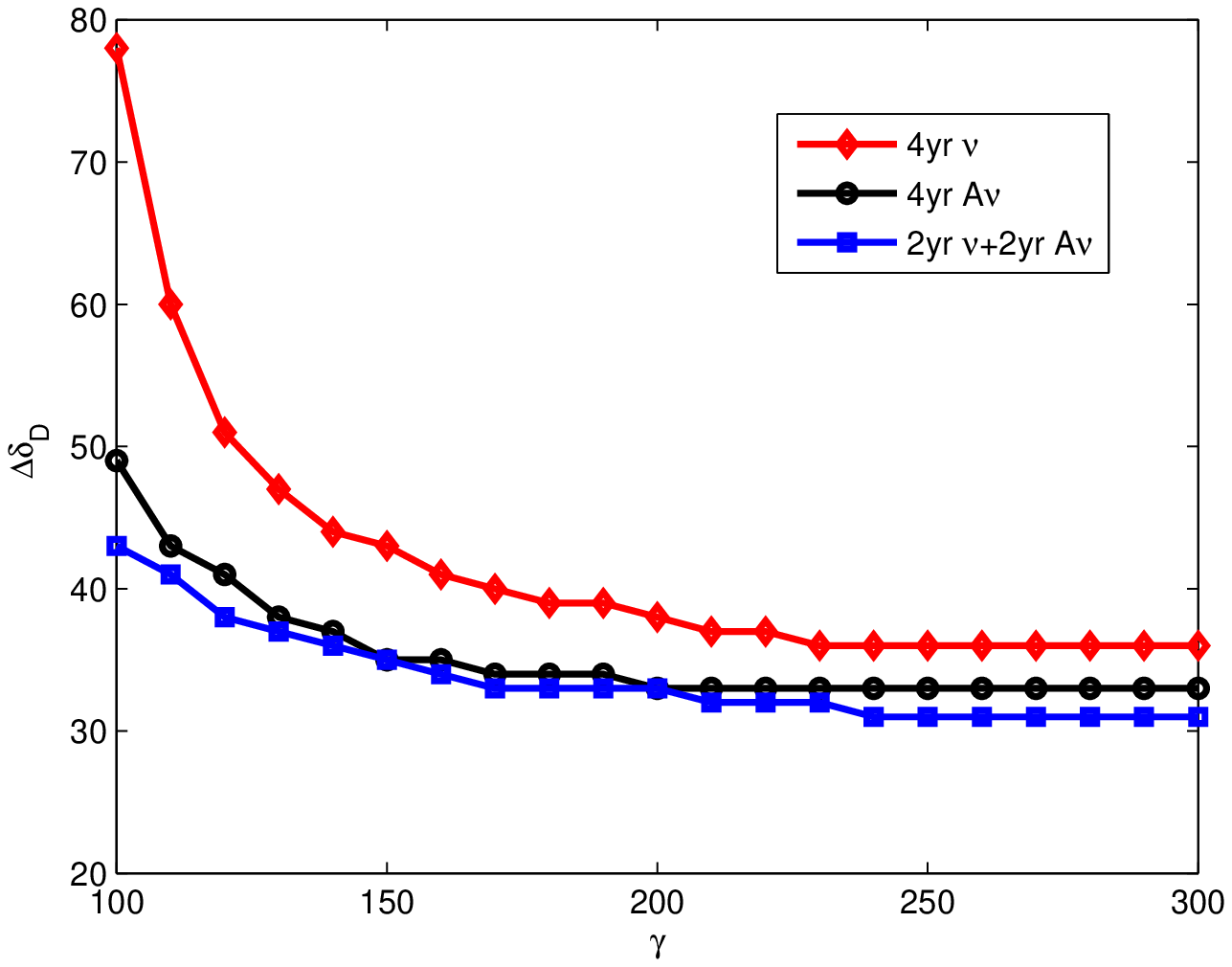}}
\subfigure[\ Inverted
hierarchy]{\includegraphics[width=0.49\textwidth]{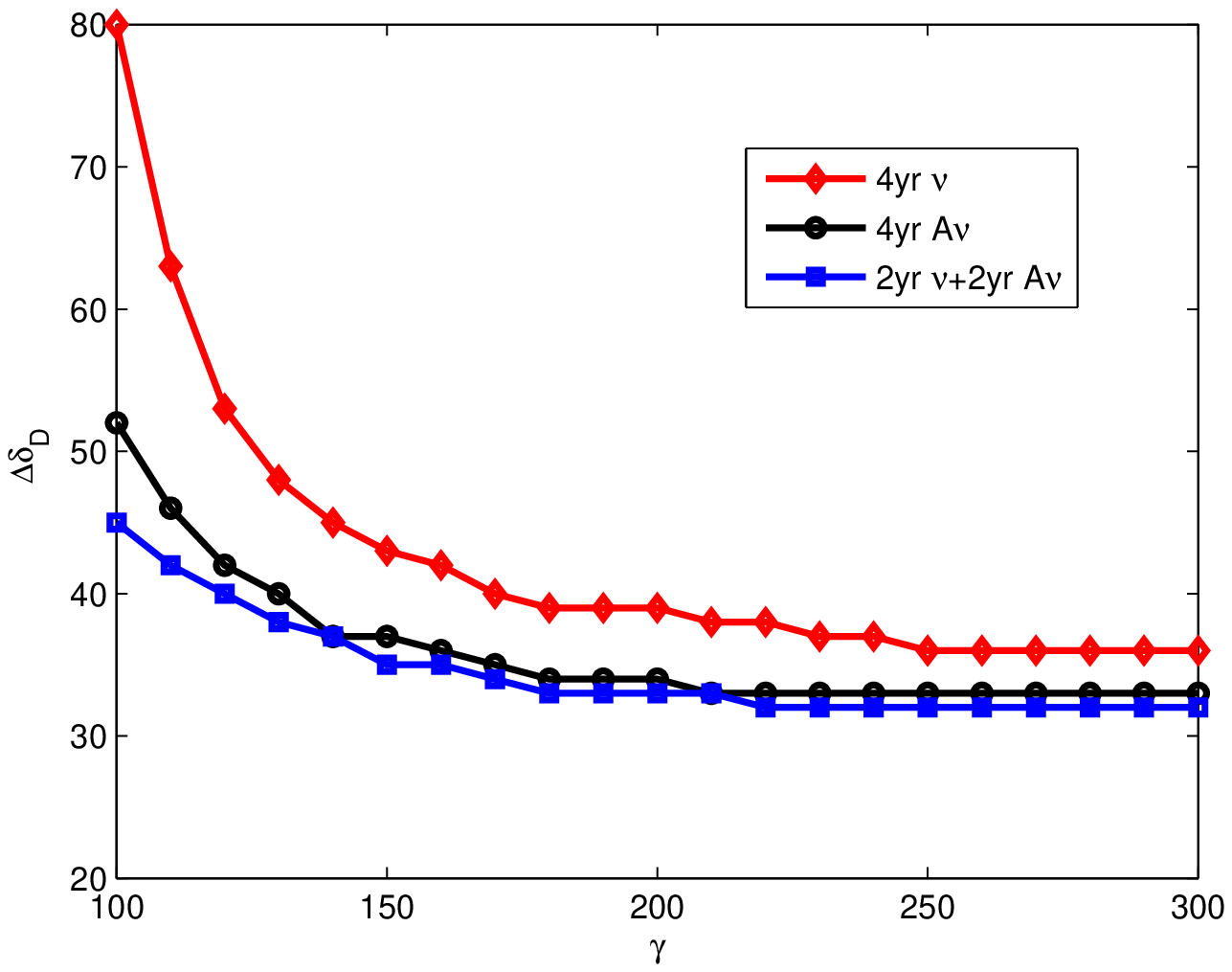}}
\subfigure[\ Normal
hierarchy]{\includegraphics[width=0.49\textwidth]{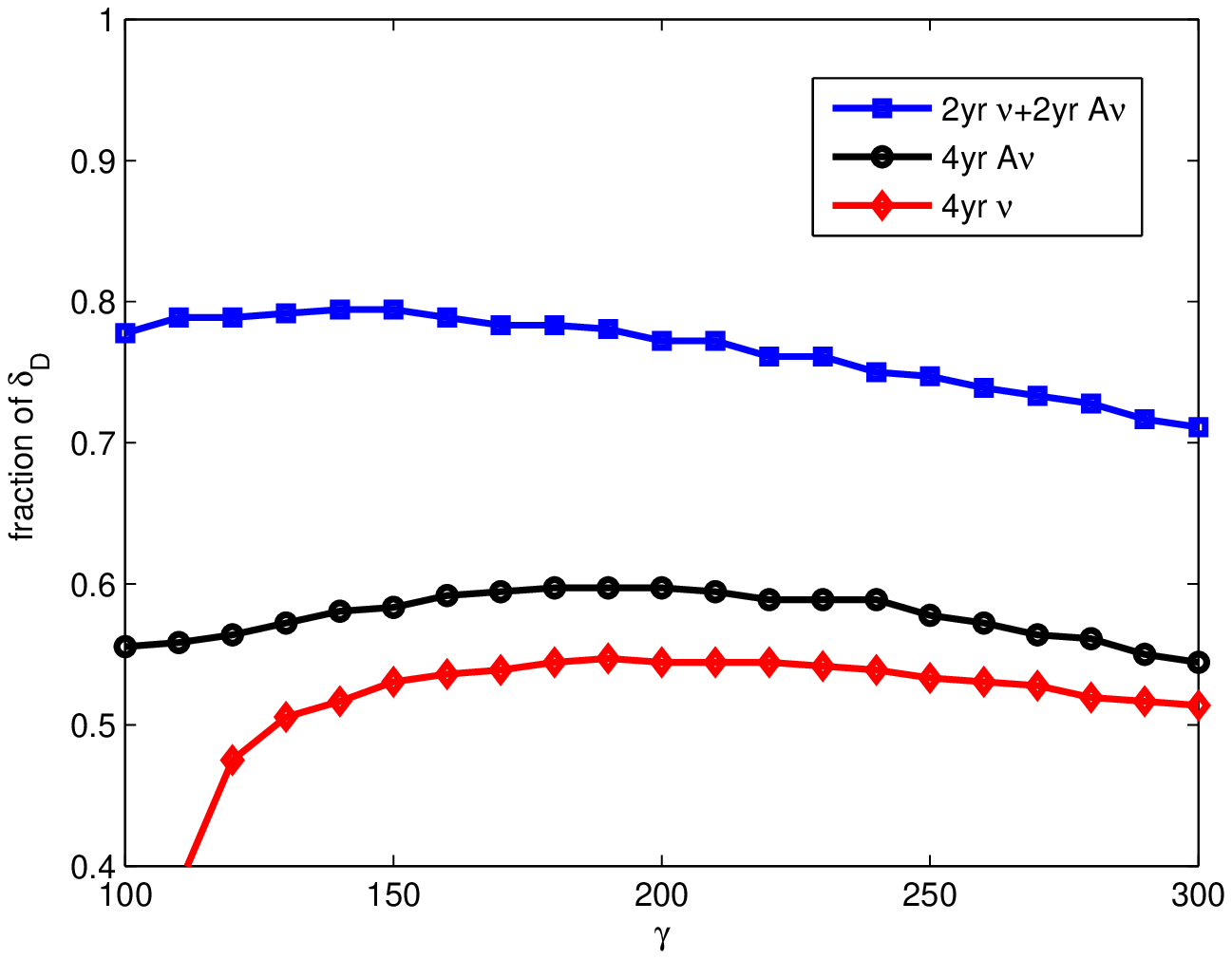}}
\subfigure[\ Inverted
hierarchy]{\includegraphics[width=0.49\textwidth]{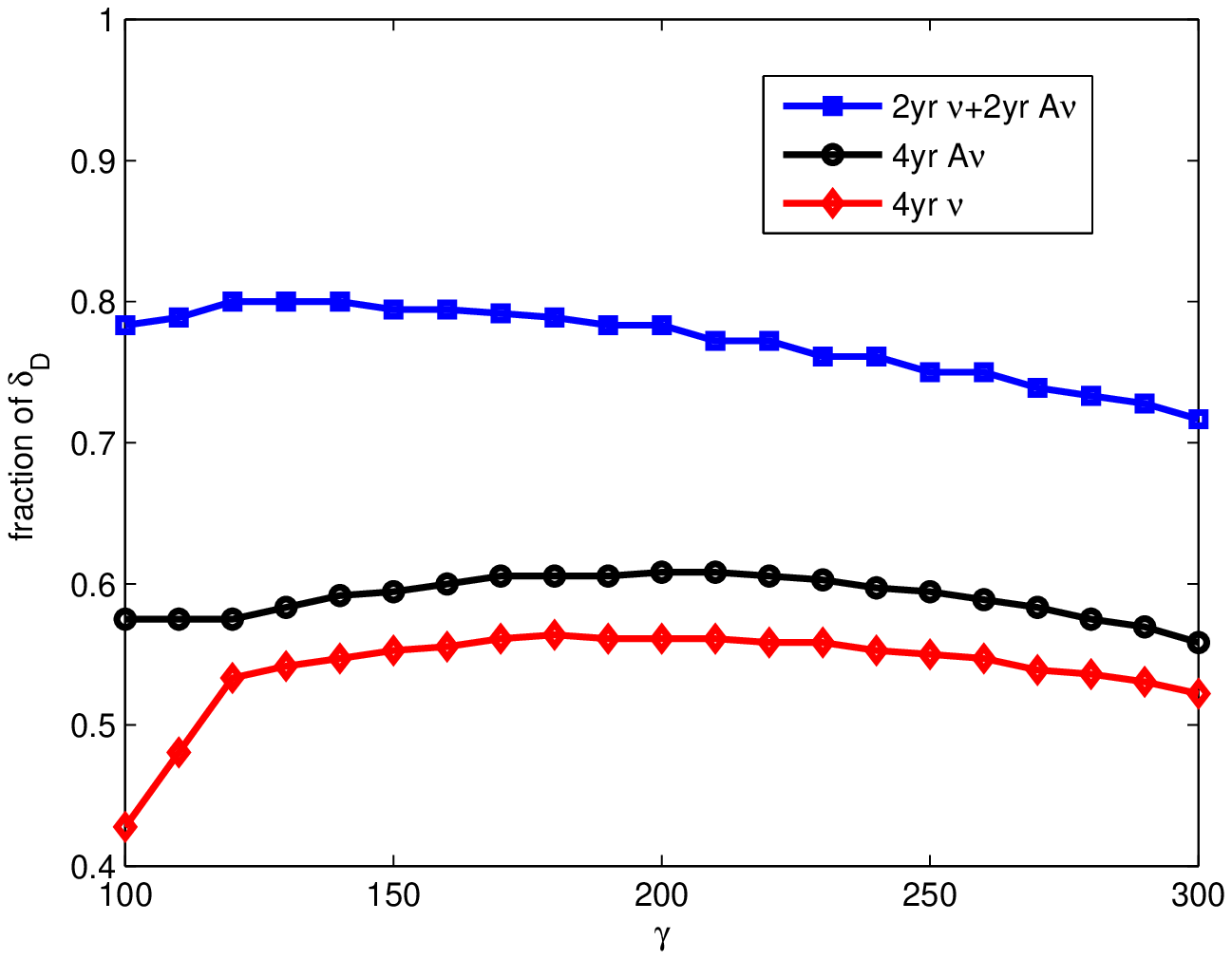}}
\end{center}
\vspace{2cm} \caption[]{The CP-discovery potential of  a 130 km
experiment (distance between CERN to Frejus) after four years of
data taking
 versus the boost factors of neutrino and antineutrino beams.
 In the left (right) panels, the hierarchy is taken to be normal (inverted).
  {\it Upper panels}: Uncertainty within which $\delta_D=90^\circ$ can be
measured  at 1$\sigma$ C.L. versus the boost factors of the $\nu_e$
and $\bar{\nu}_e$ beams from the $^{18}$Ne  and $^6$He decay. {\it
lower panels: } The fraction of the $\delta_D$ parameter for which
$CP$ can be established at higher than 95 \% C.L.}

\label{frejus}

\end{figure}

 For relatively short baselines $L\sim 100$ km, $\sin (\lambda_2-\lambda_1)/2 \ll 1$ so the contributions of the first terms in Eqs. (\ref{Pemu},\ref{PBARemu}) are subdominant relative to the second terms. As a result, the interference between the first and second terms which is the only contribution sensitive to $
 \delta_D$ will be suppressed; {\it i.e.,} when $
 \delta_D$ varies between 0 and $\pi$ the variation of $P_{e \mu}$ and $P_{\bar{e}\bar{\mu}}$ will be of order of $\sin (\lambda_2-\lambda_1)/2\sim 0.05 L/(130~{\rm km})$. On the other hand, for $L>1000$~km, $\sin  (\lambda_2-\lambda_1)/2\sim 1$ and the two terms in Eqs. (\ref{Pemu},\ref{PBARemu}) are of the same order, making the variation of the oscillation probabilities due to the variation of $
\delta_D$ of order of  the oscillation probabilities themselves. As
a result, deriving $ \delta_D$ from a 130~km setup such as CERN to
Frejus requires a different strategy than that of a very long
baseline setup with $L>1000$ km. This is demonstrated in Figs
\ref{LBNEO} and \ref{frejus}.  From Fig. \ref{LBNEO} we observe that
the pure $\bar{\nu}_e$ run in the case of the setup with $L=1300$~km
has a better prospect but as seen in Fig. 6, in case of $L=130$ km
baseline a mixed run of neutrino and antineutrino can perform better
than pure $\nu_e$ or $\bar{\nu}_e$ runs.
For $\sin (\lambda_2-\lambda_1)/2 \ll 1$, the uncertainties in
neutrino parameters (especially the uncertainties of $\theta_{13}$
and $\theta_{23}$)   induce a significant uncertainty in the
derivation of $\delta_D$. If we turn off the error in these
parameters, the performance of the CERN to Frejus setup will be
competitive with that of the LBNE setup but considering the
realistic uncertainty in these parameters as outlined in the
previous section, the sensitivity of the LBNE setup to $\delta_D$ is
much better than the $L=130$ km setup. This can be confirmed by
comparing Figs. 5 and 6.

For the 130 km setup, the oscillation probabilities can be
approximately written as $P_{\bar{e}\bar{\mu}}\simeq |i
s_{12}^mc_{23}\sin (\lambda_2-\lambda_1)+s_{13}^ms_{23}
 e^{i\delta_D}(e^{i \lambda_3}-1)|^2$ and
$P_{e\mu}\simeq |ic_{12}^mc_{23}\sin (\lambda_2-\lambda_1)+s_{13}^ms_{23}
 e^{-i\delta_D}(e^{i \lambda_3}-1)|^2$. Since we are far from the 31-resonance, $s_{13}^m$ is not much different from $s_{13}$ and is approximately the same for normal and inverted hierarchies. As a result, replacing $\delta_D\to \pi -\delta_D$ and $\Delta m_{13}^2\to -\Delta m_{13}^2$ ({\it i.e.,} $\lambda_3 \to -\lambda_3$), the oscillation probability does not change. That is why in Fig \ref{frejus}, the $
\Delta \delta_D$ plots for normal and inverted hierarchies are
practically the same. If we take a value other than $90^\circ$ as
the true value of $\delta_D$, we will not have such a symmetry.
However, as seen in Figs. \ref{LBNEO} and \ref{frejus}, the general
behavior  for normal and inverted hierarchies are similar. With the
present SPS setup, CERN cannot enhance $\gamma$ over 150 for the
$^6$He  ions \cite{Mezzetto:2003ub}. On the other hand, from Fig.
\ref{frejus}, we observe that with $L=130$ km, there is no point in
seeking higher values of $\gamma$. In fact, at $\gamma=150$, the
fraction of CP-violating $\delta_D$ parameter for which CP-violation
can be established is slightly higher than that for $\gamma>250$.

From comparing Figs. \ref{LBNEO} and \ref{frejus}, we observe that
the best performance can be achieved by $L=1300$ km setup and
antineutrino run. For example, while with CERN to Frejus setup,
$\delta_D=90^\circ$ can be measured with only uncertainty of $\Delta
\delta_D=35^\circ$; with  a 1300~km setup, the uncertainty can be
lowered down to   $\Delta \delta_D=15^\circ$.  Notice that for these
setups, the same detector (500 kton WC) is assumed. Although with
longer baselines, the flux decreases but instead
$\lambda_2-\lambda_1$ in Eq. \ref{PBARemu} becomes larger so a
moderate precision in the $P_{\bar{e}\bar{\mu}}$ measurement will
suffice to extract $\delta_D$. For measuring $\delta_D=90^\circ$,
the $L=2300$ km setup with the  $\bar{\nu}_e$ run seems to be
competitive with the $L=1300$ km setup; however, the fraction of
$\delta_D$ to be established by $L=1300$ km setup is considerably
higher. Among the setups that we have considered the $L=1300$ km
setup with a 500 kton WC and the $\bar{\nu}$ run seems to be the
most promising one.  In Fig \ref{LBNEO}, we observe that for
$200<\gamma<300$, the curves corresponding to the $\bar{\nu}_e$ run
are almost flat.

\begin{figure}
\begin{center}
\subfigure[\ Neutrino beam, normal
hierarchy]{\includegraphics[width=0.49\textwidth]{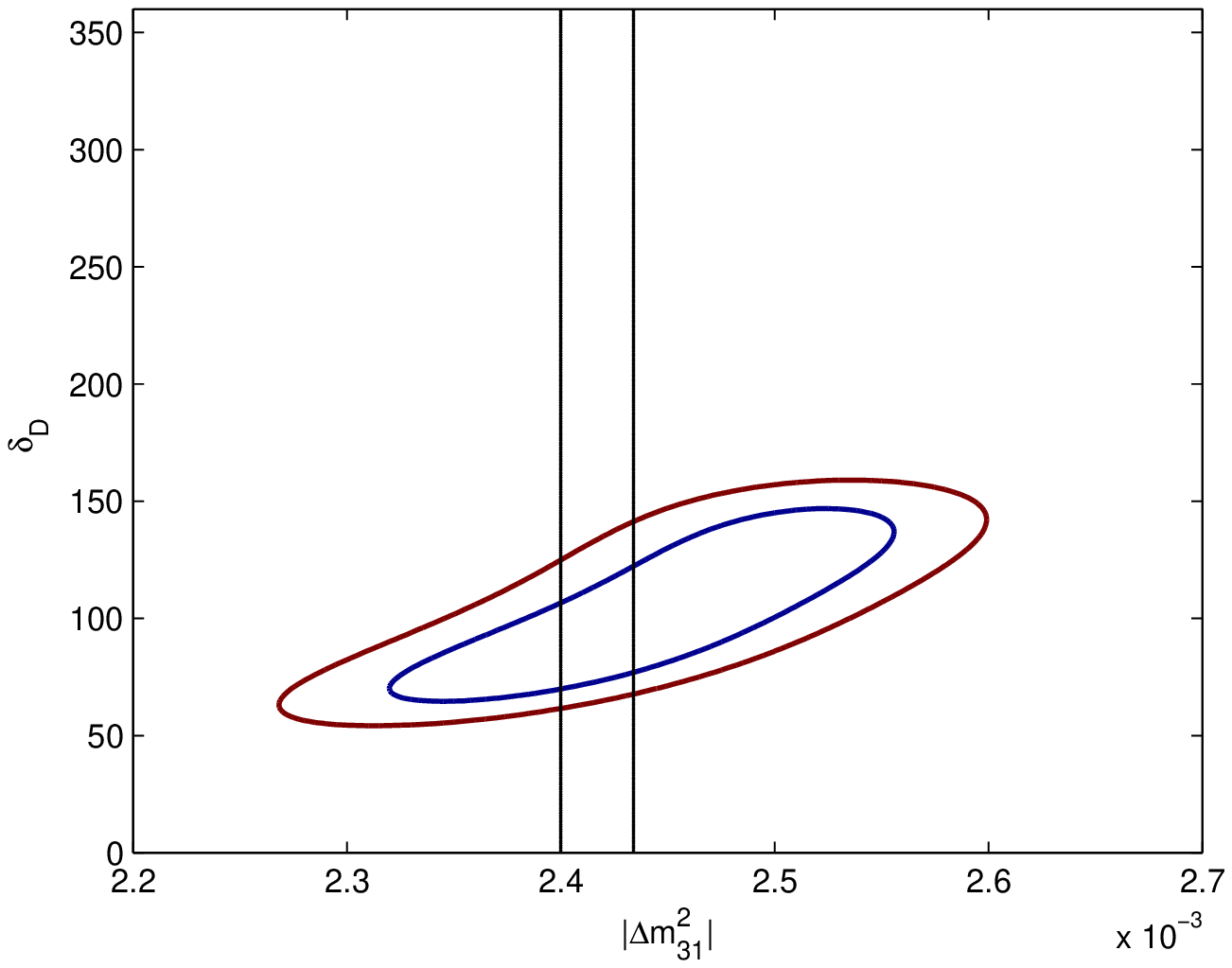}}
\subfigure[\ Neutrino beam, inverted
hierarchy]{\includegraphics[width=0.49\textwidth]{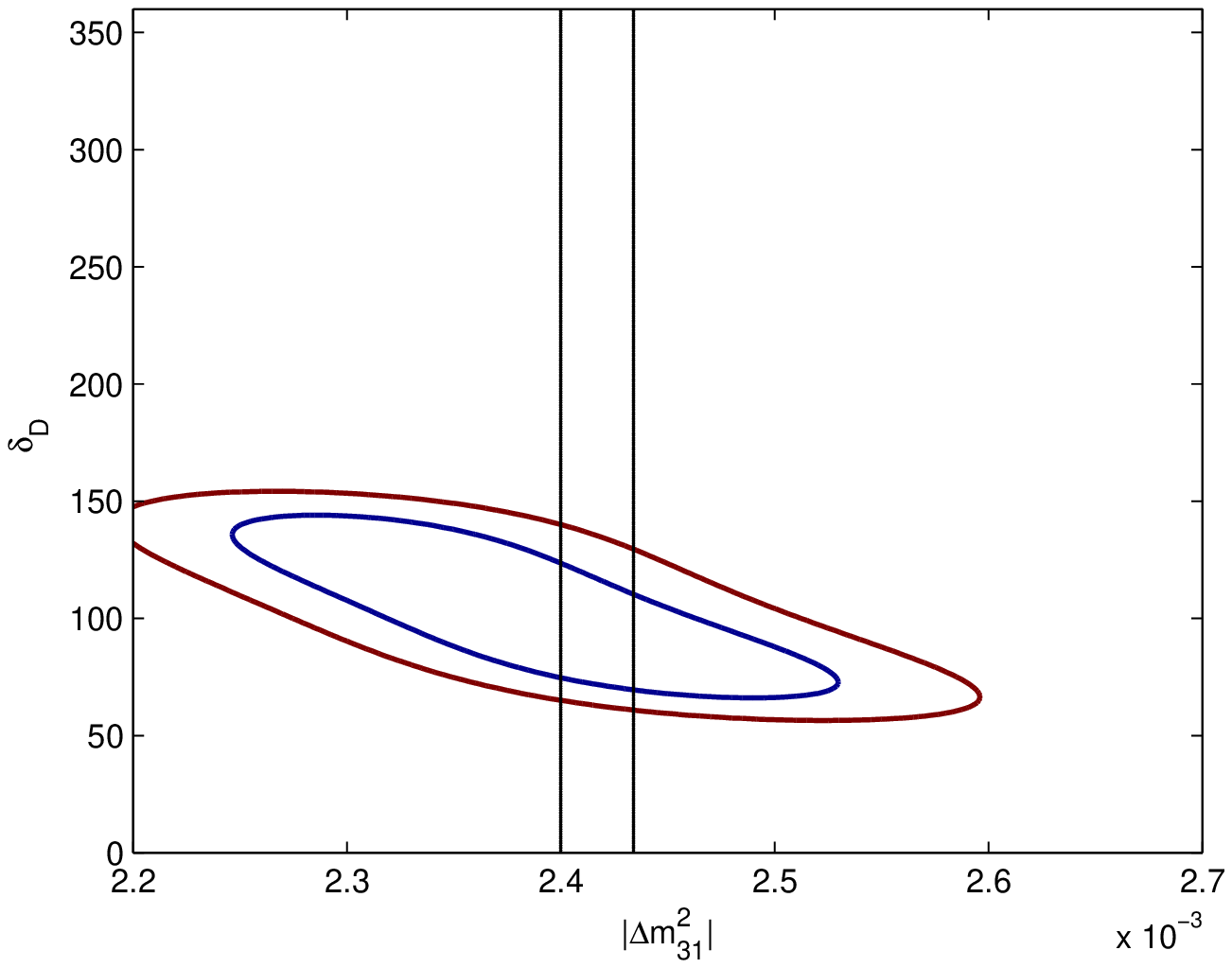}}
\subfigure[\ Antineutrino beam, normal
hierarchy]{\includegraphics[width=0.49\textwidth]{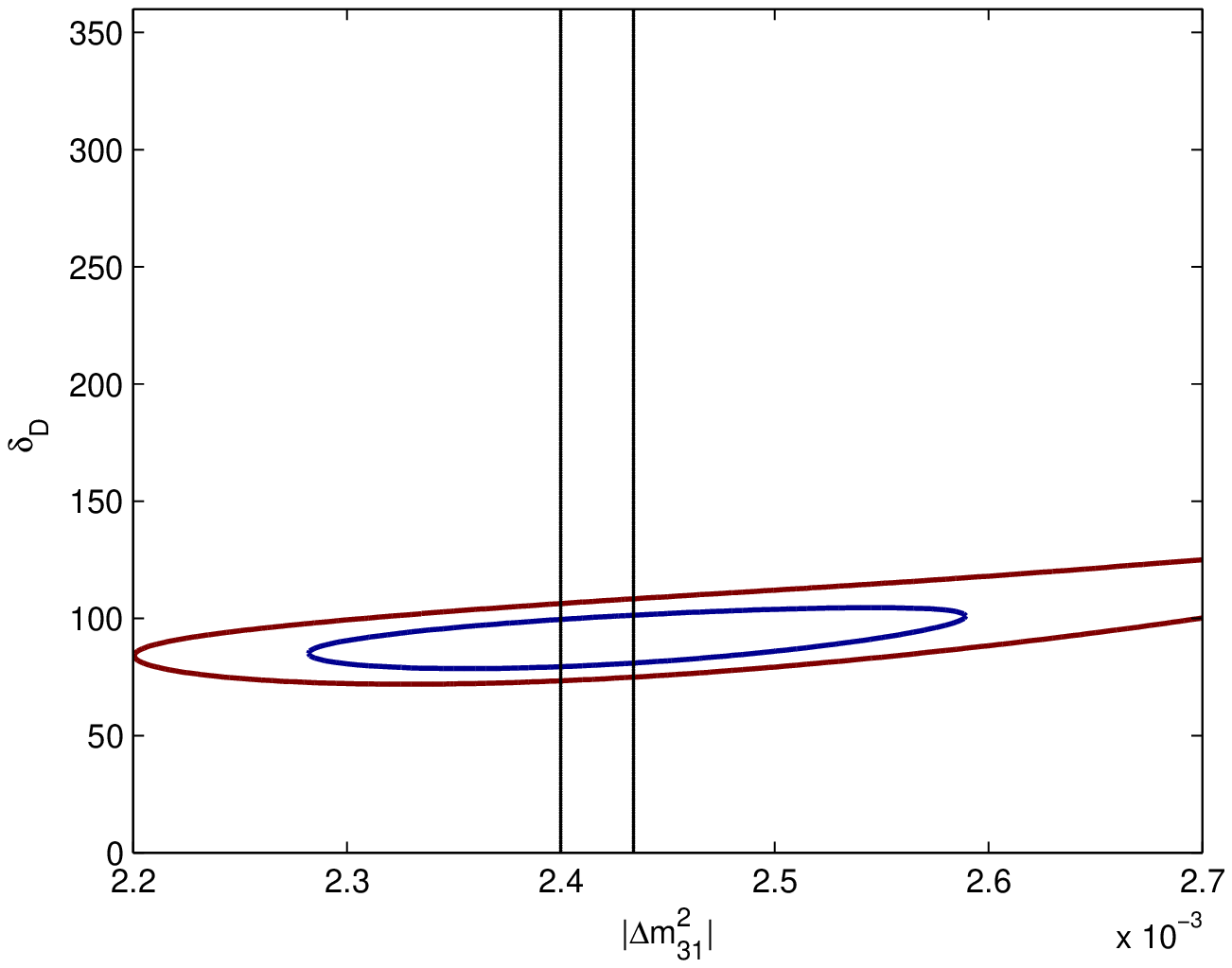}}
\subfigure[\ Antineutrino beam, inverted
hierarchy]{\includegraphics[width=0.49\textwidth]{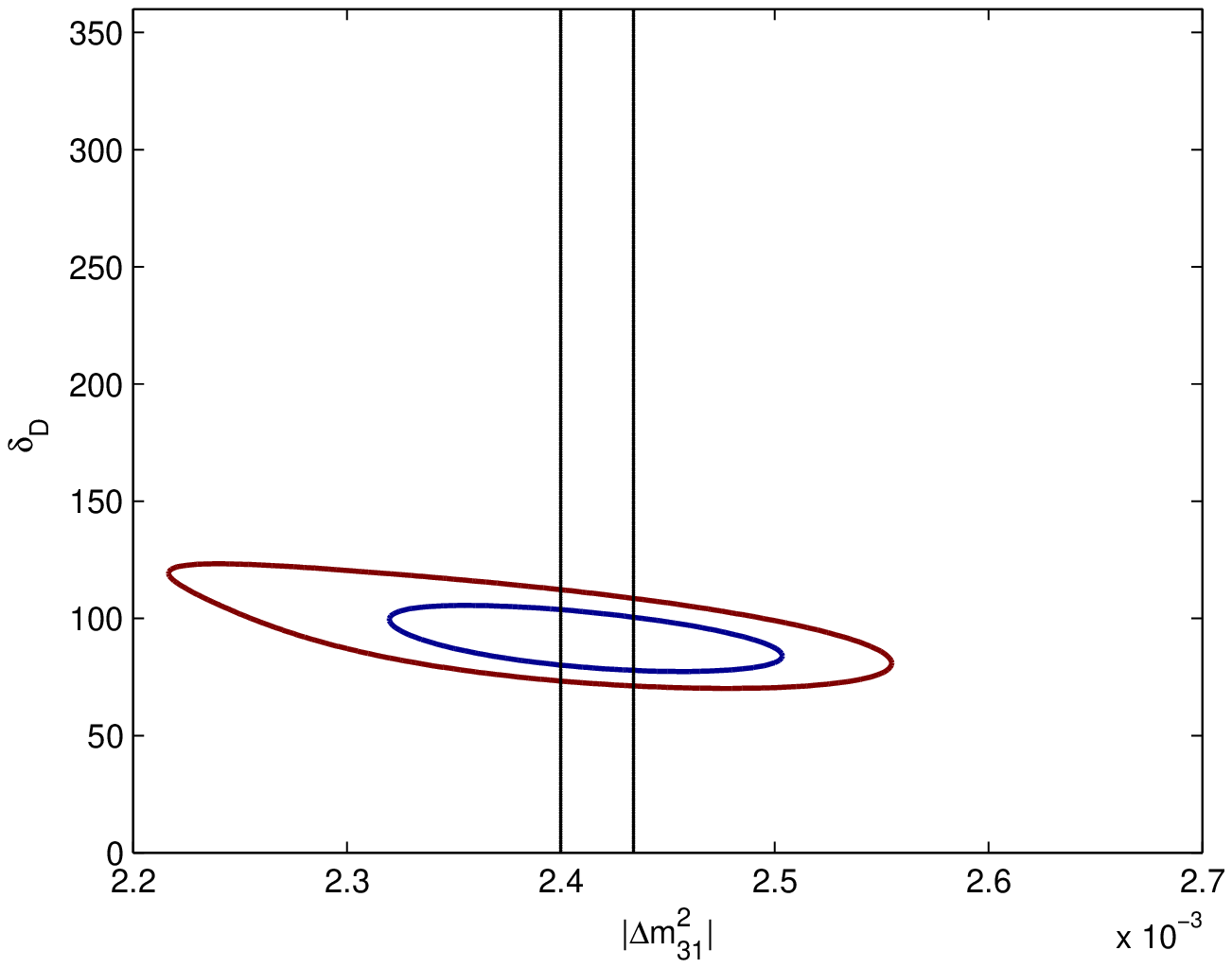}}
\end{center}
\vspace{2cm}
\caption[]{Determination of $\delta_D$ and $\Delta m_{31}^2$ with
 $\gamma=300$ and $L=1300$ km after four years of data taking.
The contours show  68 \% C.L. and 95 \% C.L. In upper (lower)
panels, neutrino (antineutrino) beam with  $2.2 \times 10^{18}$
($5.8 \times 10^{18}$) decays per year is assumed. In left (right)
panels, the hierarchy is taken to be normal (inverted). The true
value of $\delta_D$ is taken to be $90^\circ$. For normal (inverted)
hierarchy, we take  $\Delta m_{31}^2=2.421\times 10^{-3}~{\rm eV}^2$
($\Delta m_{31}^2=-2.35\times 10^{-3}~{\rm eV}^2$). The vertical
lines show a 0.7\% uncertainty in $\Delta m_{31}^2$ (e.g.,  for
normal hierarchy $\Delta m_{31}^2=2.421\times 10^{-3}~{\rm
eV}^2\times (1\pm 0.7\% )~{\rm eV}^2$).}

\label{contours}

\end{figure}

As expected for $L>1000$ km, the results are highly sensitive to the
central values of $\Delta m_{31}^2$. The oscillatory behavior in
Figs \ref{L-dd}  that we discussed before implies such a
sensitivity. In fact, the setup that we are proposing can
simultaneously extract $\delta_D$ and $\Delta m_{31}^2$. Figs.
\ref{contours} show 68 \% and 95 \% C.L. contours for $\gamma=300$
and $L=1300$ km  (LBNE). In drawing these plots, the hierarchy is
assumed to be known; however, the measured value of $\Delta
m_{31}^2$ is not used.
 The precision
in  $\Delta m_{31}^2$ can drastically be improved by forthcoming
experiments.
 In
\cite{WinteratPINGU}, it is shown that combining the T2K and PINGU
results, 0.7 \% precision in $\Delta m_{31}^2$ is achievable.  In
Fig.\ref{contours} the vertical lines show the $0.7 \%$ uncertainty
in $\Delta m_{31}^2$ around the ``true'' value of $\Delta m_{31}^2$.
As seen from the figure, for the case of the antineutrino beam the
uncertainty in $\Delta m_{31}^2$ will not significantly increase the
uncertainty in the  $ \delta_D$ determination.

\section{Conclusions}
Measuring the CP-violating phase by a beta beam facility has been extensively studied in the literature. Most of the recent studies have  focused on relatively high energy beams with $\gamma>300$. The reason is that for a given baseline, the number of detected neutrinos increases approximately as $\gamma^3$.
However, for lower energy beta beam, large volume WC detectors
\cite{economy} can be employed that can compensate for the decrease of flux and cross section. Moreover with the relatively large value of $\theta_{13}$ chosen by the nature, having enough statistics will not
be the most serious challenge for measuring the CP-violating phase. Considering these facts, we have explored the CP-discovery reach of an
intermediate  energy beta beam for various baselines and different neutrino vs antineutrino combinations using the GLoBES software \cite{Globes}. We have discussed the precision with which $\delta_D$ can be measured, assuming that by the time that the required facilities are ready the hierarchy is also determined. Our results do not depend much on which mass ordering is chosen.

 We have found  that a setup with only antineutrino run with $200<\gamma<300$ and a baseline of $L=1300$ km has an excellent discovery potential. Four years run of such a setup with $5.8\times 10^{18}$ $^6$He decays per year can establish  CP-violation at 95 \% C.L. for more than 85 \% of the $\delta_D$ parameter range. If $\delta_D=90^\circ$, this setup can determine it with impressive
precision $\delta_D=90^\circ \pm 8^\circ$ for inverted hierarchy and
$\delta_D=90^\circ \pm 7^\circ$ for normal hierarchy at 1$\sigma$
C.L. Such a baseline corresponds to the distance between FermiLAB to
Sanford underground research facility in South Dakota. A baseline of
$L=1300$ km  seems to be close to the optimal distance to measure
the Dirac CP-violating phase. We have found that for this baseline a
setup with  intermediate values of $\gamma$ in the range
$200<\gamma<300$ with a 500 kton WC detector can outperform that
with  $\gamma=450$ and 50 kton TASD.

  For the very long baselines with $L>1000$~km, a pure antineutrino source from $^6$He enjoys a better performance than a mixed neutrino antineutrino run. On the other hand for shorter baselines, a balanced neutrino antineutrino mode gives better results.
 We  have specifically discussed the CERN to Frejus setup with $L=130$ km baseline.
 We have found that with two years of neutrino mode from  $2.2 \times 10^{18}$ decays of
  $^{18}$Ne per year combined with two years of antineutrino mode
  from  $5.8 \times 10^{18}$ decays of $^6$He per year  both with
  $\gamma$=150 (the largest boost that can be obtained for $^6$He with
  the present SPS accelerator at CERN \cite{Mezzetto:2003ub}),  the  CP-violation
  can be established for about 80 \%   of the $\delta_D$ parameter range.
  With such setup and runtime, if the true value of $\delta_D$ is
  equal to $90^\circ$, it can be measured as $\delta_D=90^\circ \pm 18^\circ$ at 1$\sigma$ C.L.
 By increasing $\gamma$ to higher values, the precision in the
 $\delta_D$ measurements slightly improves however still with a similar detector
 and antineutrino run, the performance of $L=1300$ km  can be better.

 In {\it sum}, our conclusion is that a beta beam facility  with  $200<\gamma<300$, baseline of $L\simeq 1300$ km and 500 kton WC running in the antineutrino mode from $^6$He decay is an optimal option for establishing the CP-violation in the lepton sector and the measurement of $\delta_D$.
The location of source and detector can be respectively FermiLAB and
Sanford underground laboratory in south Dakota.
 \section*{Acknowledgements}
  The authors would like to thank A. Yu. Smirnov for encouragement and very useful comments. They also thank F. Terranova  and W. Winter
  for fruitful comments. P. B. acknowledges H. Mosadeq for technical help in running the computer codes.
  They also acknowledges partial support from the  European Union FP7  ITN INVISIBLES (Marie Curie Actions, PITN- GA-2011- 289442).
   Y.F. thanks the staff of Izmir technical institute (IZTECH) where a part of this work was carried out for generous support and hospitality.
   The authors also thank the anonymous referee for useful
   remarks.


%




\begin{thebibliography}{10}

\bibitem{An:2012eh}
  F.~P.~An {\it et al.}  [DAYA-BAY Collaboration],
   Phys.\ Rev.\ Lett.\  {\bf 108},
  171803 (2012)  [arXiv:1203.1669 [hep-ex]].

\bibitem{Ahn:2012nd}
  J.~K.~Ahn {\it et al.}  [RENO Collaboration],
  Phys.\ Rev.\
  Lett.\  {\bf 108}, 191802 (2012)  [arXiv:1204.0626 [hep-ex]].  
\bibitem{Abe:2011fz}
  Y.~Abe {\it et al.}  [DOUBLE-CHOOZ Collaboration],
  Phys.\ Rev.\ Lett.\  {\bf 108}, 131801 (2012)  [arXiv:1112.6353 [hep-ex]].  
 \bibitem{superbeams}
   A.~Blondel, I.~Efthymiopoulos and G.~Prior,
   J.\ Phys.\ Conf.\ Ser {\bf 408} (2013).

  \bibitem{WB}
   M.~V.~Diwan, D.~Beavis, M.~-C.~Chen, J.~Gallardo, S.~Kahn, H.~Kirk, W.~Marciano and W.~Morse {\it et al.},
  Phys.\ Rev.\ D {\bf 68}, 012002 (2003)  [hep-ph/0303081].  

\bibitem{Farzan:2002ct}
  Y.~Farzan and A.~Y.~Smirnov,
   Phys.\ Rev.\ D {\bf 65}, 113001 (2002)  [hep-ph/0201105].  

  \bibitem{triangles}
   H.~Zhang and Z.~-z.~Xing,
  Eur.\ Phys.\ J.\ C {\bf 41}, 143 (2005)  [hep-ph/0411183];  
A.~Bandyopadhyay {\it et al.}  [ISS Physics Working Group Collaboration],
   Rept.\ Prog.\ Phys.\  {\bf 72}, 106201 (2009)  [arXiv:0710.4947 [hep-ph]];
      G.~Ahuja and M.~Gupta,
  Phys.\ Rev.\ D {\bf 77}, 057301 (2008)  [hep-ph/0702129 [HEP-PH]];
   Z.~-z.~Xing and H.~Zhang,
  Phys.\ Lett.\ B {\bf 618}, 131 (2005)  [hep-ph/0503118];
 S.~Antusch, S.~F.~King, C.~Luhn and M.~Spinrath,
   Nucl.\ Phys.\ B {\bf 850}, 477 (2011)  [arXiv:1103.5930 [hep-ph]];
A.~Dueck, S.~Petcov and W.~Rodejohann,
   Phys.\ Rev.\ D {\bf 82}, 013005 (2010)  [arXiv:1006.0227 [hep-ph]];
Z.~-z.~Xing and S.~Zhou,
   Phys.\ Lett.\ B {\bf 666}, 166 (2008)  [arXiv:0804.3512 [hep-ph]].  



  \bibitem{Zucchelli}
   P.~Zucchelli,
  Phys.\ Lett.\ B {\bf 532}, 166 (2002);  
C.~Volpe,
  arXiv:0802.3352 [hep-ph];  
    C.~Volpe,
  Prog.\ Part.\ Nucl.\ Phys.\  {\bf 64}, 325 (2010)  [arXiv:0911.4314 [hep-ph]];  
   C.~Orme,
  JHEP {\bf 1007}, 049 (2010)  [arXiv:0912.2676 [hep-ph]];  
 P.~Coloma, A.~Donini, P.~Migliozzi, L.~Scotto Lavina and F.~Terranova,
  Eur.\ Phys.\ J.\ C {\bf 71}, 1674 (2011)  [arXiv:1004.3773 [hep-ph]];  
    J.~Bernabeu, C.~Espinoza, C.~Orme, S.~Palomares-Ruiz and S.~Pascoli,
   AIP Conf.\ Proc.\  {\bf 1222}, 174 (2010);  
     S.~K.~Agarwalla, A.~Raychaudhuri and A.~Samanta,
   Phys.\ Lett.\ B {\bf 629}, 33 (2005)  [hep-ph/0505015]; 
  S.~K.~Agarwalla, S.~Choubey and A.~Raychaudhuri,
  Nucl.\ Phys.\ B {\bf 798}, 124 (2008)  [arXiv:0711.1459 [hep-ph]].  
\bibitem{Sanjib}
S.~K.~Agarwalla and P.~Huber,
   Phys.\ Lett.\ B {\bf 693} (2010) 114  [arXiv:0909.2257 [hep-ph]].  

\bibitem{Orme}
D.~Meloni, O.~Mena, C.~Orme, S.~Palomares-Ruiz and S.~Pascoli,
    JHEP {\bf 0807}, 115 (2008)  [arXiv:0802.0255 [hep-ph]];  
D.~Meloni, O.~Mena, C.~Orme, S.~Pascoli and S.~Palomares-Ruiz,
  PoS NUFACT {\bf 08}, 133 (2008).  


\bibitem{Agarwalla:2013tza}
 S.~K.~Agarwalla, Y.~Kao and T.~Takeuchi,
  arXiv:1302.6773 [hep-ph].  
\bibitem{blennow} M.~Blennow and A.~Y.~Smirnov,
   Adv.\ High Energy Phys.\  {\bf 2013} (2013) 972485  [arXiv:1306.2903 [hep-ph]].  

\bibitem{memphys}
T.~R.~Edgecock, O.~Caretta, T.~Davenne, C.~Densham, M.~Fitton, D.~Kelliher, P.~Loveridge and S.~Machida {\it et al.},
   Phys.\ Rev.\ ST Accel.\ Beams {\bf 16} (2013) 021002  [arXiv:1305.4067 [physics.acc-ph]].  
  \bibitem{Huber}
  P.~Huber, M.~Lindner, M.~Rolinec and W.~Winter,
  Phys.\ Rev.\ D {\bf 73}, 053002 (2006)  [hep-ph/0506237].  

\bibitem{Mezzetto:2003ub}
  M.~Mezzetto,
   J.\ Phys.\ G {\bf 29} (2003) 1771  [hep-ex/0302007].  
\bibitem{economy}
 W.~Winter,
   Phys.\ Rev.\ D {\bf 78}, 037101 (2008)  [arXiv:0804.4000 [hep-ph]].  

\bibitem{Greenfield}
S.~K.~Agarwalla, S.~Choubey, A.~Raychaudhuri and W.~Winter,
   JHEP {\bf 0806}, 090 (2008)  [arXiv:0802.3621 [hep-ex]].  



\bibitem{Globes}
P.~Huber, M.~Lindner and W.~Winter,
  Comput.\ Phys.\ Commun.\  {\bf 167}, 195 (2005)  [hep-ph/0407333];  
P.~Huber, J.~Kopp, M.~Lindner, M.~Rolinec and W.~Winter,
  Comput.\ Phys.\ Commun.\  {\bf 177}, 432 (2007)  [hep-ph/0701187];  
http://www.mpi-hd.mpg.de/personalhomes/globes.
\bibitem{NuFit} M.~C.~Gonzalez-Garcia, M.~Maltoni,
J.~Salvado and T.~Schwetz,
   JHEP {\bf 1212} (2012) 123  [arXiv:1209.3023 [hep-ph]];  
{\it see also,} v1.1 results in www.nu-fit.org.
\bibitem{pingu}
E.~K.~Akhmedov, S.~Razzaque and A.~Yu.~Smirnov,
   JHEP {\bf 02} (2013) 082   [arXiv:1205.7071 [hep-ph]];  
  S.~K.~Agarwalla, T.~Li, O.~Mena and S.~Palomares-Ruiz,
  arXiv:1212.2238 [hep-ph];  
D.~Franco {\it et al.},
   JHEP {\bf 1304} (2013) 008  [arXiv:1301.4332 [hep-ex]];  
 T.~Ohlsson, H.~Zhang and S.~Zhou,
   Phys.\ Rev.\ D {\bf 88} (2013) 013001
  [arXiv:1303.6130 [hep-ph]].;  
A.~Esmaili and A.~Yu.~Smirnov,
  JHEP {\bf 1306} (2013) 026
  [arXiv:1304.1042 [hep-ph]];  
 M.~Ribordy and A.~Y.~Smirnov,
 Phys.\ Rev.\ D {\bf 87} (2013) 113007
  [arXiv:1303.0758 [hep-ph]].  
 \bibitem{WinteratPINGU}
W.~Winter,
  Phys.\ Rev.\ D {\bf 88} (2013) 013013
  [arXiv:1305.5539 [hep-ph]].  
 \bibitem{DayaPINGU}
  M.~Blennow and T.~Schwetz,
   JHEP {\bf 1309} (2013) 089
  [arXiv:1306.3988 [hep-ph]].  

  \bibitem{octant}
S.~K.~Agarwalla, S.~Prakash and S.~U.~Sankar,
  JHEP {\bf 1307} (2013) 131
  [arXiv:1301.2574 [hep-ph]].  

\bibitem{Capozzi:2013psa}
  F.~Capozzi, E.~Lisi and A.~Marrone,
  arXiv:1309.1638 [hep-ph].  

\bibitem{Tortola:2012te}
  D.~V.~Forero, M.~Tortola and J.~W.~F.~Valle,
   Phys.\ Rev.\ D {\bf 86} (2012) 073012  [arXiv:1205.4018 [hep-ph]].  
\bibitem{Raut:2012dm}
  S.~K.~Raut,
    Mod.\ Phys.\ Lett.\ A {\bf 28} (2013) 1350093  [arXiv:1209.5658 [hep-ph]].  
\bibitem{prem} A. M. Dziewonski and D. L. Anderson, Preliminary reference earth model, Phys. Earth Planet Interiors \textbf{25} (1981) 297; F. F. Stancy, Physics of the Earth, 2nd ed., Wiley, 1977.

\bibitem{Parke}
A.~Jansson, O.~Mena, S.~J.~Parke and N.~Saoulidou,
  Phys.\ Rev.\ D {\bf 78} (2008) 053002  [arXiv:0711.1075 [hep-ph]].  

\bibitem{29}
 P.~Huber, M.~Lindner and W.~Winter,
   Nucl.\ Phys.\ B {\bf 645} (2002) 3  [hep-ph/0204352].  

\bibitem{Agostino}
L.~Agostino {\it et al.}  [MEMPHYS Collaboration],
   JCAP {\bf 1301} (2013) 024  [arXiv:1206.6665 [hep-ex]].  

\bibitem{CS}
M.~D.~Messier,
   UMI-99-23965;  
E.~A.~Paschos and J.~Y.~Yu,
   Phys.\ Rev.\ D {\bf 65} (2002) 033002  [hep-ph/0107261].  

\bibitem{AguilarArevalo:2013hm}
  A.~A.~Aguilar-Arevalo {\it et al.}  [MiniBooNE Collaboration],
  Phys.\ Rev.\ D {\bf 88} (2013) 032001
   [arXiv:1301.7067 [hep-ex]].

   \bibitem{migration}
   P.~Huber, M.~Lindner and W.~Winter,
  Comput.\ Phys.\ Commun.\  {\bf 167} (2005) 195  [hep-ph/0407333];  
P.~Huber, J.~Kopp, M.~Lindner, M.~Rolinec and W.~Winter,
  Comput.\ Phys.\ Commun.\  {\bf 177} (2007) 432  [hep-ph/0701187].  




 \bibitem{LBNE}
http://lbne.fnal.gov/

\bibitem{LBNO}
  A.~Stahl, C.~Wiebusch, A.~M.~Guler, M.~Kamiscioglu, R.~Sever, A.~U.~Yilmazer, C.~Gunes and D.~Yilmaz {\it et al.},
  CERN-SPSC-2012-021.  




\end{thebibliography}
\end{document}